\documentclass[a4paper,11pt]{article}
\usepackage[T1]{fontenc}
\usepackage[utf8]{inputenc}
\usepackage{lmodern}
\usepackage[margin=2.5cm]{geometry}
\usepackage{amsmath,amssymb,amsthm}
\usepackage{graphicx}
\usepackage{booktabs,longtable}
\usepackage{xcolor}
\usepackage{listings}
\usepackage{footnote}
\usepackage{microtype}
\usepackage{setspace}
\usepackage[authoryear,round]{natbib}
\usepackage[hidelinks]{hyperref}
\graphicspath{{Fig/}{./}}
\makesavenoteenv{tabular}
\makesavenoteenv{table}

\theoremstyle{plain}

\theoremstyle{remark}

\theoremstyle{definition}

\defcitealias{Istat2021}{ISTAT (2022)}
\defcitealias{Istat2022}{ISTAT et al.\ (2023)}

\title{A two-step model to study inclusivity in Italian early childhood education and care services}
\author{\href{https://orcid.org/0000-0002-1141-3041}{Angela Andreella}\thanks{Corresponding author: \href{mailto:angela.andreella@unive.it}{angela.andreella@unive.it}.}
\and Federico Caldura
\and \href{https://orcid.org/0000-0002-1099-530X}{Gaia Bertarelli}
\and \href{https://orcid.org/0000-0002-7113-4458}{Stefano Campostrini}}
\date{\small Department of Economics, Ca' Foscari University of Venice\\
\small Cannaregio, 873, 30100 Venice, Italy}

\begin{document}
\maketitle

\begin{abstract}
This study defines and measures inclusivity in Italian early childhood education and care (ECEC) services, highlighting the gap between legislative principles and local realities. Although Decree No. 65/2017 establishes inclusivity as a guiding principle of ECEC provision, prioritizing universal access, evidence on its effective application across local contexts remains limited. We propose a two-step model to examine this misalignment: first, a latent trait model estimates an inclusivity index; second, a mixed-effects quantile regression model analyzes its distribution across Italian regions. We additionally implement a small area estimation strategy to derive reliable indicators at the Nomenclature of Territorial Units for Statistics 3 level (Italian provinces), a territorial scale finer than the survey’s effective domain but often relevant for decision-making on these services. We observe substantial variability both between and within regions, pointing to the lack of a cohesive and inclusive ECEC system at the national and regional levels. This paper urges policymakers to standardize ECEC access, offering policy-relevant insights and recommendations to improve inclusivity in Italian ECEC services.
\end{abstract}

\noindent\textbf{Keywords:} ECEC services, inclusivity, latent trait model, mixed-effects quantile regression, small area estimation.

\section{Introduction}
	Early childhood education and care (ECEC) services are crucial in cultivating well-rounded economic and social conditions within countries \citep{plantenga2009provision}. A substantial body of evidence shows that access to ECEC services contributes to higher female labor market participation and greater gender equality in employment \citep{landivar2021states, addabbo2012allocation, del2002effect, chiuri2000quality, meyers1999public}, while also affecting demographic behavior, including fertility decisions \citep{luci2013impact, haan2011can, del2002effect}. In addition, high-quality ECEC provision is associated with improved cognitive and developmental outcomes for children, particularly in the early years \citep{felfe2018does, van2018children, brilli2016does, felfe2015can}.

	In Italy, ECEC services were formally introduced in 1971 as ``social services of public interest''. From the late 1990s onward, additional forms of provision were progressively established across the country. Despite these developments, overall coverage remains limited, falling below the European benchmark of 33\%. Moreover, provision is unevenly distributed across the territory, resulting in marked regional disparities. Specifically, the availability of ECEC services is often concentrated in large cities and northern regions, leaving many geographic areas with limited access to such services (ISTAT et al., \citeyear{Istat2022}). The unequal distribution makes it difficult for several families to meet their childcare needs and underscores the need for targeted strategies to ensure fair and consistent access nationwide. Addressing these disparities is crucial for filling the current gaps in ECEC coverage in Italy \citep{andreella2024spatial, andreella2023suitable}. 
	
	Beyond documenting the scarcity and uneven distribution of ECEC services across the Italian territory, a pivotal aspect for policymakers is understanding how much of this limited and variable supply is actually accessible to all the subgroups of the population. An increase in the number of available places does not necessarily translate into broader access if structural, economic, or sociocultural barriers persist. In this respect, the notion of inclusivity becomes central. The Italian Legislative Decree No. 65/2017 defines ECEC services as inclusive if they are ``open to all children, respecting the individuality, culture, and religion of the child and the socioeconomic status''. While this provides a normative benchmark, its empirical operationalization remains underspecified. Preliminary evidence provided by \citetalias{Istat2022} points to marked territorial heterogeneity in how inclusivity principles are implemented, reflecting differences in governance, organizational arrangements, and local practices. This heterogeneity is closely related to the uneven geography of public provision, more prevalent in the North and in larger cities, which matters because most services offering financial support for socio-economically disadvantaged families (e.g., low-income households, foreign children, and children with disabilities) are public (ISTAT et al., \citeyear{Istat2022}). Overall, local variation in management models, fee reductions, admission criteria, and disability accommodations suggests that inclusivity is not yet a harmonized, system-wide feature of Italian ECEC services and may depend more on where children live than on their needs.
	
	Therefore, expanding ECEC provision should go beyond increasing the number of available places (coverage) and ensure that services are accessible and inclusive for families regardless of their geographic, economic, and social conditions. Assessing inclusivity requires reliable and comparable measures that capture its multiple dimensions, i.e., social, economic, and territorial. Without such measurement, policy interventions may inadvertently reinforce existing territorial and socioeconomic disparities rather than reduce them. Yet the effective accessibility of ECEC services is often overlooked in research and policymaking \citep{vandenbroeck2014accessibility}, especially for disadvantaged children from low-income families \citep{sylva2007family, fram2008race, ghysels2011unequal, oecdstst}. Addressing this gap calls for new data and appropriate statistical methods that can synthesize multidimensional information while identifying which service features are associated with higher or lower inclusivity.
	
	The literature on constructing an inclusivity index in the context of ECEC is limited. Existing contributions have primarily examined inclusivity in labor market settings \citep{el2021inclusivity}, while studies in the educational domain have largely focused on higher education institutions rather than on ECEC services. For example, \citet{jaegler2022measure} proposes a formative measurement model to analyze inclusivity in higher education, considering the perspectives of four stakeholders: students, alumni, employees, and external partners. Inclusivity is defined across four dimensions for each stakeholder group: gender, identity, disability, and social openness. Similarly, \citet{faura2022building} develop a composite index to assess inclusivity in universities based on five areas: resources, measures, and actions supporting access and retention; physical accessibility in university premises; technological and information accessibility; resources for support and adaptations in the teaching-learning and assessment process; and other programs/actions. In contrast, studies addressing inclusivity in ECEC, such as those by \citet{carrington2014learning} and \citet{duran2005primeras}, often rely on the index for inclusion proposed by \citet{booth2002index} and \citet{vaughan2002index}. This framework evaluates inclusivity along three dimensions: culture, policies, and practices. It is defined through 45 indicators and approximately 500 questions, allowing all school stakeholders, i.e., teachers, administrators, parents, governors, and students, to contribute data via questionnaires.
	
	This paper adopts a reflective measurement strategy. In contrast to the formative index proposed by \citet{booth2002index}, we conceptualize inclusivity as a latent construct that is reflected in a set of observed indicators. The indicators are treated as manifestations of the same underlying trait and are therefore expected to covary. This choice is consistent with both the structure of the available data and a view of inclusivity as a multidimensional yet coherent concept. A formative specification would instead require a comprehensive and non-overlapping set of causal indicators, an assumption that is difficult to sustain given data constraints and the strong correlations among the observed variables. Reflective latent variable models are widely used to measure complex social constructs, including well-being and deprivation \citep{moretti2021small, gasparini2013multidimensional, luzzi2008cluster}, and provide a statistically grounded framework for synthesizing correlated information into an index.
	
	We then propose a two-step procedure to define and analyze inclusivity in ECEC services. Consistent with the legislative definition and given the structure of the available survey data (ISTAT et al., \citeyear{Istat2022}), we conceptualize inclusivity along two main dimensions: a social dimension, capturing access for children with disabilities and migrant backgrounds, and an economic dimension, reflecting fee structures and tuition reduction mechanisms that affect affordability. We first construct an inclusivity index using a constrained configural multiple-group two-dimensional (social and economic) latent trait model \citep{mirt, moustaki2000generalized, darrell1970fitting}, with ownership (private and public) as the grouping variable. The same prespecified allocation of indicators to the social and economic dimensions is used for public and private services, while item intercepts and discrimination parameters are allowed to vary by ownership. This distinction is relevant because public and private providers may implement inclusivity through different mechanisms, particularly with respect to access rules, fee structures, and economic-support measures (ISTAT et al., \citeyear{Istat2022}). Because no item parameters are constrained to be equal across groups, however, the resulting latent scales are ownership-specific and do not support direct comparisons of absolute inclusivity levels between public and private providers. In the second stage, we examine the distribution of the estimated index through a quantile regression model with regional random effects \citep{geraci2007quantile, geraci2014linear}. This approach allows us to analyze territorial heterogeneity beyond mean differences and to account explicitly for within-region variability in inclusivity levels. Finally, the survey design permits reliable direct estimation only at the macro-area level (North, Center, and South/Islands). However, given the central role of regional and sub-regional governance in ECEC provision, this level of aggregation is insufficient for policy-relevant analysis. We therefore implement a Small Area Estimation (SAE) strategy \citep{rao2015small} using a modified version of the Empirical Best Prediction (EBP) approach of \citet{molina2010small} called census Empirical Best Prediction (census SAE EBP) proposed by \citet{guadarrama2016comparison}, to obtain statistically reliable estimates of the inclusivity index at the Nomenclature of Territorial Units for Statistics (NUTS) 3 level (Italian province), explicitly accounting for the uncertainty from the latent trait estimation, following the approach proposed by  \citet{moretti2021small}. The empirical results reveal substantial heterogeneity across and within regions, highlighting the territorially differentiated nature of inclusivity in the Italian ECEC system.
	
	The remainder of the paper is organized as follows. Section~\ref{data} describes the data. Section~\ref{inclusivity} outlines the methodological framework, including the latent trait model, the random-effects quantile regression, and the SAE procedure. Section~\ref{ltm_results} reports the empirical findings. Section~\ref{discussion} concludes with a discussion of the results and directions for further research.

	\section{Data}\label{data}
	
	The novel data used to explore the concept of inclusivity in Italian ECEC services are from an Italian probabilistic sample survey on ECEC services, conducted yearly by the Italian National Institute of Statistics (ISTAT), in collaboration with Ca' Foscari University of Venice and the Department for Family Policies (ISTAT et al., \citeyear{Istat2022}). The survey collects information such as families' demand, accommodation capacity, occupancy rates, and quality of ECEC services. To the best of our knowledge, it represents the only official nationwide source providing micro-level information on ECEC services through direct interviews with service providers. 
	
	We use data from the 2021/2022 survey wave (ISTAT et al., \citeyear{Istat2022}), which employed a stratified sampling design with equal selection probabilities within each stratum. The initial sampling weights, defined as the inverse of the inclusion probabilities within strata, were adjusted through post-stratification using the updated ISTAT register for the 2020/2021 educational year (ISTAT, \citeyear{Istat2021}). Stratification is defined by region, service type (nursery and ``sezione primavera''\footnote{The ``sezione primavera'' is an educational project designed for children between $24$ and $36$ months of age.}), and ownership (public and private). The main estimation domains are macro-areas (North, Center, and South/Islands), crossed with service type and ownership; sample allocation follows an optimal allocation procedure with respect to these domains and is then distributed proportionally across regions to ensure territorial coverage. Within each stratum, units are selected by simple random sampling. The planned gross sample size was about 3000 units (approximately 25\% larger than the theoretical minimum) to account for expected non-response (ISTAT et al., \citeyear{Istat2022}). The distribution of the population and the selected sample by macro-area, service type, and ownership is reported in Table~\ref{tab:sampling}. The final responding sample comprises 1374 ECEC services. After excluding units with missing information on key items, the analytical sample includes 1323 services, covering 104 provinces across 19 regions.
	
	\begin{table}[!t]
		\centering
		\caption{Distribution of ECEC services in the population, selected sample, and responding sample by macro-area, service type, and ownership.}
		\label{tab:sampling}
		\begin{tabular*}{.9\columnwidth}{@{\extracolsep\fill}l|rrr@{\extracolsep\fill}}
			\textbf{Domain} &\textbf{Population} & \textbf{Selected sample} & \textbf{Responding sample} \\
			\toprule
			North & 6207 & 941 & 536 \\
			Central & 2707 & 813 & 375 \\
			South/Islands & 2764 & 1260 & 463 \\
			Nursery & 8957 & 1565 & 750 \\
			``Sezione primavera'' & 2721 & 1449 & 624 \\
			Private & 7327 & 1974 & 819 \\
			Public & 4351 & 1040 & 555 \\
			\midrule
			Italy & 11678 & 3014 & 1374 \\
			\bottomrule
		\end{tabular*}
	\end{table}
	
	In this paper, we focus on the survey section capturing inclusivity as accessibility to ECEC services. Based on the available items, we operationalize inclusivity along two dimensions (social and economic) consistent with the literature \citep{plantenga2009provision}. While inclusivity can also encompass cultural, linguistic, and other barriers, these dimensions are not observed in the questionnaire; our measure should therefore be interpreted as a data-constrained but empirically grounded operationalization. Given the heterogeneity of conceptual frameworks in the literature \citep{booth2002index, faura2022building, jaegler2022measure}, we adopt a theory-guided specification constrained by the available survey items. In the next section, we model inclusivity as a latent construct within a reflective measurement framework, interpreting selected indicators as manifestations of an underlying trait.
	
	The variables used in our analysis are described in Table~\ref{tab:var}. As mentioned above, we want to define inclusivity by exploring the social (i.e., the first $2$ variables) and economic (i.e., the last $8$ variables) aspects of Italian ECEC services, as measured in the quoted survey. Exploratory plots of these variables are presented in Appendix~\ref{app:descriptive}.
	
	Three indicators are constructed as follows to consider the actual presence of foreigners in the territory. Let $c_k$ denote the number of foreign children enrolled in the ECEC service $k$, where $k \in \{1, \dots, 1323\}$ indexes the surveyed ECEC services, and let $f_p$, where $p \in \{1, \dots, 104\}$, denote the number of resident foreigners aged $0$--$2$ years in province $p$ (NUTS~3 level). The foreign child enrollment rate $F_p$ for the province $p$ is then defined as 
	\begin{equation*}
		F_p = \dfrac{\sum_{k \in \mathcal{P}_p} c_k}{f_p}
	\end{equation*}
	where $\mathcal{P}_p$ is the set containing the indices of the ECEC services in the province $p$. The foreign variable at level $\lambda \in \{0.25,0.5,0.75\}$ is then defined as:
	\begin{equation}\label{eq:foreign}
		\text{Foreign}_k(\lambda) = 
		\begin{cases}
			1 & \text{if } F_p \ge Q_{F}(\lambda)\\
			0 & \text{otherwise} 
		\end{cases} \quad \forall k \in \mathcal{P}_p
	\end{equation}
	where $Q_{F}(\lambda)$ is the quantile of $F = \{F_1, \dots, F_{104}\}$ at level $\lambda$. To represent the information encoded by these nested indicators as a single ordinal variable, we consider
	\begin{equation*}
		\text{Foreign}_k
		=
		\sum_{\lambda\in \{0.25,\,0.50,\,0.75\}}
		\text{Foreign}_k(\lambda).
	\end{equation*}
	Thus, the variable $\text{Foreign}_k\in\{0,1,2,3\}$. Higher values indicate a higher estimated province-level participation of foreign children in ECEC services. Hence, $\text{Foreign}_k$ is constant for all services within the same province. All remaining variables in Table~\ref{tab:var} are instead defined at the service level $k$.

	\begin{table}[!t]
		\centering
		\caption{Description of the variables used in the analysis from the Italian ECEC services sample survey (ISTAT et al., \citeyear{Istat2022}).  \label{tab:var}}%
		\begin{tabular*}{\columnwidth}{@{\extracolsep\fill}llll@{\extracolsep\fill}}
			\textbf{Variable} & \textbf{Description} \\
			\toprule
			Foreign &
			Ordinal variable equal to $0$ if $F_p<Q_F(0.25)$;
			$1$ if $Q_F(0.25)\leq F_p<Q_F(0.50)$;\\
			&$2$ if $Q_F(0.50)\leq F_p<Q_F(0.75)$; and
			$3$ if $F_p\geq Q_F(0.75)$ \\
			& where $Q_F(\lambda)$ is the quantile of the foreign child enrollment rate at level $\lambda$.\\
			Disability & $1$ if there is at least one child with a disability enrolled, $0$ otherwise\\
			\midrule
			Meal & $1$ if the meal is included in the fee, $0$ otherwise\\
			Fee & $1$ if there is no entry fee, $0$ otherwise\\
			Disability fee & $1$ if there is a reduction for a child with a disability, $0$ otherwise\\
			Full fee & $1$ if full tuition exemption is granted,  $0$ otherwise\\
			ISEE \footnote{ISEE is a tool used in Italy to assess the economic situation of individuals and families based on various factors such as income, assets, and family composition.} & $1$ if there is a reduction for ISEE,  $0$ otherwise \\
			Child & $1$ if there is a reduction for other children  (not necessarily enrolled), $0$ otherwise \\
			Social Services & $1$ if there is a reduction for social services  indications, $0$ otherwise \\
			Family & $1$ if there is a reduction for another family  condition, $0$ otherwise \\
			\bottomrule
		\end{tabular*}
	\end{table}

	\section{Inclusivity index}\label{inclusivity}
	
	This section describes the models employed. Subsection~\ref{ltm} delineates the latent trait model \citep{moustaki2000generalized,darrell1970fitting} that is used to measure inclusivity. Subsection~\ref{ebp} defines the SAE model \citep{molina2010small} used to obtain reliable estimates of the inclusivity index at the NUTS 3 level. Finally, Subsection~\ref{qr} shows the mixed-effects quantile regression model \citep{spagnolo2020use, geraci2014linear}, which permits exploring the differences in inclusivity between the Italian regions. Survey weights are incorporated in both modeling stages, albeit at different levels of aggregation. The latent trait model is estimated using service-level sampling weights to correct for unequal inclusion probabilities in the survey design. By contrast, the quantile regression employs regional weights to account for territorial representation in the distributional analysis. As a robustness check, Appendix \ref{app:ltm} reports the latent trait model estimated without sampling weights and a test for the informativeness of sampling weights  \citep{asparouhov2007testing}.
	
	\subsection{Latent trait model}\label{ltm}
	
	We define below the latent trait model \citep{moustaki2000generalized,andersen1983latent}, which is widely used in several fields such as education \citep{hambleton1977latent}, epidemiology \citep{duncan1986utility}, and binary multi-item analysis in general \citep{darrell1970fitting}. Here, the latent trait model is employed to estimate the concept of inclusivity using the variables defined in Table~\ref{tab:var}. 
	
	A central specification choice is whether measurement is formative or reflective \citep{coltman2008formative}. Formative models treat indicators as defining the construct, typically via an aggregation rule with externally assigned weights. Reflective models instead treat the indicators as manifestations of an underlying latent trait that accounts for their dependence. We adopt a reflective specification for both conceptual and empirical reasons. Conceptually, more inclusive services should be more likely to exhibit favorable social and economic features. Empirically, the indicators are positively correlated, consistent with a common latent source of variation. Within the prespecified two-dimensional structure,
	the reflective framework allows item discrimination
	parameters and the association between the latent
	dimensions to be estimated from the data rather than imposed through ad hoc weighting, and it avoids the mechanical redefinition of the construct that arises in formative indices when indicators are added or omitted. Overall, the reflective latent trait model provides a statistically grounded basis for constructing an inclusivity index from correlated variables.
	
	Building on the conceptual framework outlined above, we represent inclusivity through two prespecified latent dimensions. We measure the social dimension with the ordinal \textit{Foreign} indicator and the \textit{Disability} indicator, while the economic dimension is measured by the eight fee- and support-related indicators listed in Table \ref{tab:var}. We adopt a constrained confirmatory specification in which each indicator loads only on its theoretically assigned dimension. This structure preserves the intended distinction between social and economic inclusivity and prevents sample-specific associations among the indicators from determining the factor interpretation.
	
	The model is estimated within a configural multiple-group framework \citep{mirt}, using ownership (private and public) as the grouping variable. The same pattern of fixed and freely estimated loadings is imposed for public and private services, while item intercepts, thresholds, and discrimination parameters are allowed to vary across groups. The term configural therefore refers to the measurement model's common organization and does not imply equal measurement parameters or a common latent metric. This flexibility accommodates differences in how inclusion-related practices are implemented across ownership types, particularly in relation to admission rules, fee structures, and economic-support mechanisms (ISTAT et al., \citeyear{Istat2022}). Appendix \ref{app:ltm} reports model-fit comparisons and measurement invariance analysis.
	
	Let $\mathcal{B}$ denote the index set of the nine binary indicators reported in Table~\ref{tab:var}. For service $k$ in ownership group $g\in\{1,2\}$, define $\pi_{ikg}=\Pr(X_{ikg}=1\mid\mathbf{z}_{kg})$ and assume
	\begin{equation*}
		X_{ikg}\mid\mathbf{z}_{kg}
		\sim\operatorname{Bernoulli}(\pi_{ikg}),
		\qquad i\in\mathcal{B}.
	\end{equation*}
	Let $Y_{kg}\in\{0,1,2,3\}$ denote the ordinal \textit{Foreign} indicator. Inclusivity is represented by two latent dimensions, $\mathbf{z}_{kg}=(z_{1kg},z_{2kg})^\top$, corresponding to the social and economic components, respectively. Accordingly, let $d(i)=1$ for \textit{Disability} and $d(i)=2$ for all other binary indicators. Within each ownership group, we assume $\mathbf{z}_{kg}\sim\mathcal{N}_{2}(\mathbf{0},\boldsymbol{\Sigma}_{g})$, where $\boldsymbol{\Sigma}_{g}$ has unit diagonal and group-specific off-diagonal element $\rho_g$. Thus, both latent dimensions have mean zero and variance one, while their correlation is estimated separately by ownership group. Conditional on $\mathbf{z}_{kg}$, all binary indicators and the ordinal \textit{Foreign} indicator are assumed to be mutually independent.
	
	For the binary indicators, the latent trait model is
	\begin{equation}
		\label{model}
		\log\left(\frac{\pi_{ikg}}{1-\pi_{ikg}}\right)
		=
		\beta_{0ig}
		+
		\beta_{1ig}z_{d(i),kg},
	\end{equation}
	where $\beta_{0ig}$ is the group-specific intercept and $\beta_{1ig}$ is the group-specific discrimination parameter.
	
	For the ordinal \textit{Foreign} indicator, we specify the graded-response model
	\begin{equation}
		\label{model_foreign}
		\log\left[
		\frac{\Pr(Y_{kg}\geq r\mid\mathbf{z}_{kg})}
		{\Pr(Y_{kg}<r\mid\mathbf{z}_{kg})}
		\right]
		=
		\beta_{0rg}^{F}
		+
		\beta_{1g}^{F}z_{1kg},
		\qquad r\in\{1,2,3\},
	\end{equation}
	where $\beta_{0rg}^{F}$ are group-specific cumulative intercepts satisfying
	$\beta_{01g}^{F}>\beta_{02g}^{F}>\beta_{03g}^{F}$,
	and $\beta_{1g}^{F}$ is the group-specific discrimination parameter,
	common across the three cumulative logits. At $z_{1kg}=0$, $\beta_{0rg}^{F}$ represents the log-odds of being in category $r$ or above rather than below $r$. A one-unit increase in $z_{1kg}$ multiplies these cumulative odds by $\exp(\beta_{1g}^{F})$. The common discrimination parameter across the three cumulative logits corresponds to the proportional-odds assumption of the graded-response model.

	Sampling weights at the ECEC level are incorporated into the likelihood-based estimation of $\beta_{0ig}$, $\beta_{1ig}$, $\beta_{0rg}^F$ and $\beta_{1g}^{F}$ \citep{asparouhov2005sampling}. Parameters are estimated by marginal maximum likelihood via the EM algorithm with Gauss--Hermite quadrature, as proposed by \citet{bock1988full}.  Implementation is performed using the \texttt{mirt} package in \texttt{R} \citep{mirt}.
	
	\subsection{Small area estimation model}\label{ebp}
	SAE comprises a class of statistical methods aimed at estimating population characteristics for geographic or
	demographic domains in which the available sample information is insufficient to obtain direct estimates with adequate precision
	\citep{rao2015small}. The term ``small'' therefore refers to the amount of information available for estimation rather than necessarily to the geographic extent or population size of a domain. Such domains may be defined geographically or through cross-classifications of territorial, socioeconomic, and demographic characteristics. Model-based SAE methods are commonly distinguished into unit-level approaches, which use auxiliary information for individual population units, and area-level approaches, which combine direct estimates with domain-level auxiliary information.
	
	In the ECEC survey, NUTS~3 provinces are unplanned estimation domains: the sampling design does not ensure adequate precision at this territorial level. Consequently, provincial sample sizes may be small, and some province-by-ownership domains may contain no sampled services. Direct estimates of provincial means may therefore be imprecise or
	unavailable. To address this limitation, we adopt a unit-level SAE approach based on the nested-error regression model of
	\citet{battese1988error}, hereafter referred to as the Battese--Harter--Fuller (BHF) model.
	
	Auxiliary information is obtained from the register of ECEC services. The register provides unit-level covariates describing service type (nursery or ``sezione primavera'') and geographic macro-region (North-East, North-West, Center, and South/Islands), together with province and ownership identifiers. These variables are available for both sampled services and the population of services covered by the register. They allow the model to relate observed inclusivity scores to service characteristics and to account for the composition of each provincial population when constructing predictions.
	
	We first construct a single index from the two latent dimensions estimated in Section \ref{ltm}. For service $k$ in ownership
	group $g\in\{1,2\}$, we define
	\begin{equation}\label{eq:composite_index}
		\hat{z}_{kg}
		=
		\frac{1}{2}\hat{z}_{1kg}
		+
		\frac{1}{2}\hat{z}_{2kg},
	\end{equation}
	where $\hat{z}_{1kg}$ and $\hat{z}_{2kg}$ are the latent scores estimated under the models defined by
	Equations~\eqref{model}--\eqref{model_foreign}. The composite score $\hat{z}_{kg}$ is then a summary of the two latent dimensions, rather than an additional latent factor. For policy-oriented interpretation, we use equal weights in Equation \eqref{eq:composite_index}, reflecting a normative choice to assign equal importance to the social and economic dimensions. For ease of interpretation, we rescale \eqref{eq:composite_index} using the minimum and maximum observed within each ownership group $g$. Let $\mathcal{K}_g$ denote the sampled services in group $g$. The normalized index is
	\begin{equation}\label{eq:scaledltm}
		\tilde{z}_{kg}
		=
		\frac{
			\hat{z}_{kg}
			-
			\displaystyle\min_{\ell\in\mathcal{K}_g}\hat{z}_{\ell g}
		}{
			\displaystyle\max_{\ell\in\mathcal{K}_g}\hat{z}_{\ell g}
			-
			\displaystyle\min_{\ell\in\mathcal{K}_g}\hat{z}_{\ell g}
		}.
	\end{equation}
	$\tilde{z}_{kg}$ then describes the position of service $k$ within the observed score range of its ownership group. This normalization does not establish a common metric across ownership groups; differences between public and private services should therefore not be interpreted as absolute differences in inclusivity.
	
	Let $\mathcal{U}_{pg}$ denote the register population of services belonging to province $p$ and ownership group $g$, with $N_{pg}=|\mathcal{U}_{pg}|>0$, where $|\cdot|$ denotes set cardinality. Our target is the finite-population mean of the score on the corresponding ownership-specific scale, i.e.,
	\begin{equation*}\label{eq:sae_target}
		\mu_{pg}
		=
		\frac{1}{|\mathcal{U}_{pg}|}
		\sum_{k\in \mathcal{U}_{pg}}\tilde{z}_{kpg}.
	\end{equation*}
	The scores entering the definition of $\mu_{pg}$ use the fitted ownership-specific measurement model \eqref{model}-\eqref{model_foreign} and the sample-based normalization \eqref{eq:scaledltm}. However, they can be calculated only for sampled services, whose item responses are observed. The population mean $\mu_{pg}$ must therefore be predicted. For this purpose, we fit a separate BHF model for each group $g$:
	\begin{equation}\label{eq:bhf}
		\tilde{z}_{kpg}
		=
		\mathbf{r}_{kpg}^{\top}\boldsymbol{\zeta}_g
		+
		s_{pg}
		+
		\varepsilon_{kpg},
	\end{equation}
	where $\mathbf{r}_{kpg}$ contains an intercept and indicator variables for service type and geographic macro-region. The vector $\boldsymbol{\zeta}_g$ contains the corresponding regression coefficients. The random components satisfy $s_{pg}\overset{\mathrm{iid}}{\sim} \mathcal{N}(0,\sigma_{s,g}^{2})$, $\varepsilon_{kpg}\overset{\mathrm{iid}}{\sim}
	\mathcal{N}(0,\sigma_{\varepsilon,g}^{2})$, and are mutually independent. The province-specific random intercept
	$s_{pg}$ captures residual between-province heterogeneity, whereas $\varepsilon_{kpg}$ represents service-level variability. The Gaussian model is used as a working model for the normalized score; no further transformation of the response is applied.
	
	Estimation is performed using the \texttt{R} package \texttt{emdi} \citep{kreutzmann2019emdi}. Variance components are estimated by restricted maximum likelihood (REML), with the fixed effects estimated under the fitted mixed model. The SAE model is fitted without sampling weights. This is distinct from any weighting used to estimate the first-stage measurement model.
	
	Provincial indicators are obtained using the census SAE EBP approach \citep{guadarrama2016comparison},
	a modification of the EBP framework of \citet{molina2010small}. In contrast to the finite-population construction that retains observed responses for sampled units and predicts only nonsampled responses, the census SAE EBP generates
	model-based responses for the entire register population. It therefore does not require sampled services to be individually
	linked to their corresponding register records.
	
	For the Gaussian model used here, the provincial mean predictor
	has the analytical expression
	\begin{equation}\label{eq:census_ebp_mean}
		\hat{\mu}_{pg}^{\mathrm{EBP},\infty}
		=
		\bar{\mathbf{r}}_{\mathcal{U},pg}^{\top}
		\hat{\boldsymbol{\zeta}}_g
		+
		\hat{s}_{pg},
		\qquad
		\bar{\mathbf{r}}_{\mathcal{U},pg}
		=
		\frac{1}{N_{pg}}
		\sum_{k\in \mathcal{U}_{pg}}\mathbf{r}_{kpg}.
	\end{equation}
	The superscript \(\infty\) denotes the analytical limit as the number of Monte Carlo replications tends to infinity. The first term uses the known composition of the provincial register population. The second adjusts this prediction using
	the observed scores of sampled services in that province.
	Specifically, let $\mathcal K_{pg}$ denote the set of sampled services
	in province $p$ and ownership group $g$, with
	$n_{pg}=|\mathcal K_{pg}|>0$. Then,
	\[
	\hat s_{pg}
	=
	\hat\gamma_{pg}
	\frac{1}{n_{pg}}
	\sum_{k\in\mathcal K_{pg}}
	\left(
	\tilde z_{kpg}
	-
	\mathbf r_{kpg}^{\top}\hat{\boldsymbol\zeta}_g
	\right),
	\qquad
	\hat\gamma_{pg}
	=
	\frac{
		\hat\sigma_{s,g}^{2}
	}{
		\hat\sigma_{s,g}^{2}
		+
		\hat\sigma_{\varepsilon,g}^{2}/n_{pg}
	}.
	\]
	Thus, the provincial adjustment is a shrunken average residual. Provinces with less sample information receive greater shrinkage
	toward the prediction based on register covariates. For domains without sampled services, $\hat s_{pg}=0$. In the implementation, the predictor \eqref{eq:census_ebp_mean} is approximated by Monte Carlo simulation. For each replication, responses are generated for all registered services from the fitted predictive distribution, conditional on the observed sample scores. The provincial means are then averaged across replications.
	
	Finally, uncertainty is assessed through a separate bootstrap procedure. Since the response in Equation~\eqref{eq:bhf}
	is itself estimated, the procedure also accounts for the measurement stage.  Adapting the general strategy of \citet{moretti2021small}, we re-estimate the measurement model defined in Section \ref{ltm}, reconstruct the composite scores in Equation~\eqref{eq:composite_index}, repeat the normalization in Equation~\eqref{eq:scaledltm}, and refit the BHF model in Equation~\eqref{eq:bhf}. Nonparametric resampling at the measurement stage is combined with a parametric bootstrap at the SAE stage. The resulting MSE estimates are intended to incorporate uncertainty from both stages. The complete algorithm is reported in Appendix~\ref{app:mse-inclusivity}.

	\subsection{Random-effects $\tau$-quantile regression}\label{qr}
	
	Given the skewness and differences in shape of the index distributions shown in Figure~\ref{fig:hist}, together with the
	variability observed both between and within regions in Figures~\ref{fig:median_private}--\ref{fig:median_public}, we analyze the normalized index $\tilde z_{kg}$ defined in Equation~\eqref{eq:scaledltm} using quantile regression with region-specific random intercepts. Specifically, for each ownership group $g\in\{1,2\}$, we fit linear quantile mixed models at $\tau\in\{0.25,0.50,0.75\}$ \citep{koenker2005quantile, koenker1978regression, geraci2007quantile, geraci2014linear}. This approach allows us to examine regional differences at the lower quartile, median, and upper quartile, while accounting for the clustering of services within regions \citep{spagnolo2020use, tzavidis2016longitudinal}.
	
	The models are fitted separately for public and private services because the configural measurement model does not establish
	a common latent metric across ownership groups, and the subsequent within-group rescaling does not provide such comparability. The analysis therefore focuses on regional variation within each ownership group. This approach also allows regional patterns to differ between public and private providers, reflecting their distinct institutional and regulatory settings, particularly regarding access rules, fees, and economic-support measures (ISTAT et al., \citeyear{Istat2022}).
	
	In particular, we use the approach of \citet{geraci2007quantile, geraci2014linear}, who propose the linear quantile mixed model (LQMM). Indeed, \citet{spagnolo2020use} also introduce an $M$-quantile mixed model that incorporates weights in the estimation process. Since results from both approaches were consistent (see Appendix \ref{app:lqmm_mqmm_comparison}), we opted for the first method due to the availability of a well-documented and publicly accessible \texttt{R} package, which facilitates reproducibility. Estimation is then performed using the \texttt{lqmm} \texttt{R} package \citep{lqmm}.
	
	Let $\tilde{z}_{kjg}$ denote the ownership-specific inclusivity index defined in Equation~\eqref{eq:scaledltm} for service $k$ located in region $j$ and belonging to ownership group $g\in\{1,2\}$, where $j \in \{1,\ldots,J\}$ and $J=19$. For each ownership group and quantile level $\tau\in\{0.25,0.50,0.75\}$, we specify the linear quantile mixed model 
	\begin{equation}
		\label{eq:quantile_model}
		Q_{\tilde{z}_{kjg}}(\tau\mid u_{jg,\tau})
		=
		\gamma_{0g,\tau}+u_{jg,\tau},
	\end{equation}
	where $\gamma_{0g,\tau}$ is the fixed intercept, representing the conditional $\tau$-quantile when $u_{jg,\tau}=0$, and $u_{jg,\tau}$ is the region-specific random deviation. Thus, $\gamma_{0g,\tau}+u_{jg,\tau}$ represents the conditional
	$\tau$-quantile of inclusivity for services in region $j$ and ownership group $g$.
	
	For each ownership group and quantile level, we assume that the region-specific random intercepts are independent across regions, with $u_{jg,\tau}\sim\mathcal{N}(0,\sigma_{u,g,\tau}^{2})$. The random-effect variance is therefore allowed to differ across ownership groups and quantile levels. Finally, owner-specific weights are inserted in the definition of the log-likelihood function. We considered the proportions of the Italian ECEC services in region $j$ as regional weights. Therefore, the weights used in the model are defined at the second level, i.e., the regional level.
	
	\section{Results}\label{ltm_results}
	
	\subsection{Latent inclusivity index}
	
\begin{figure}[!t]
		\centering
		\includegraphics[width=\columnwidth]{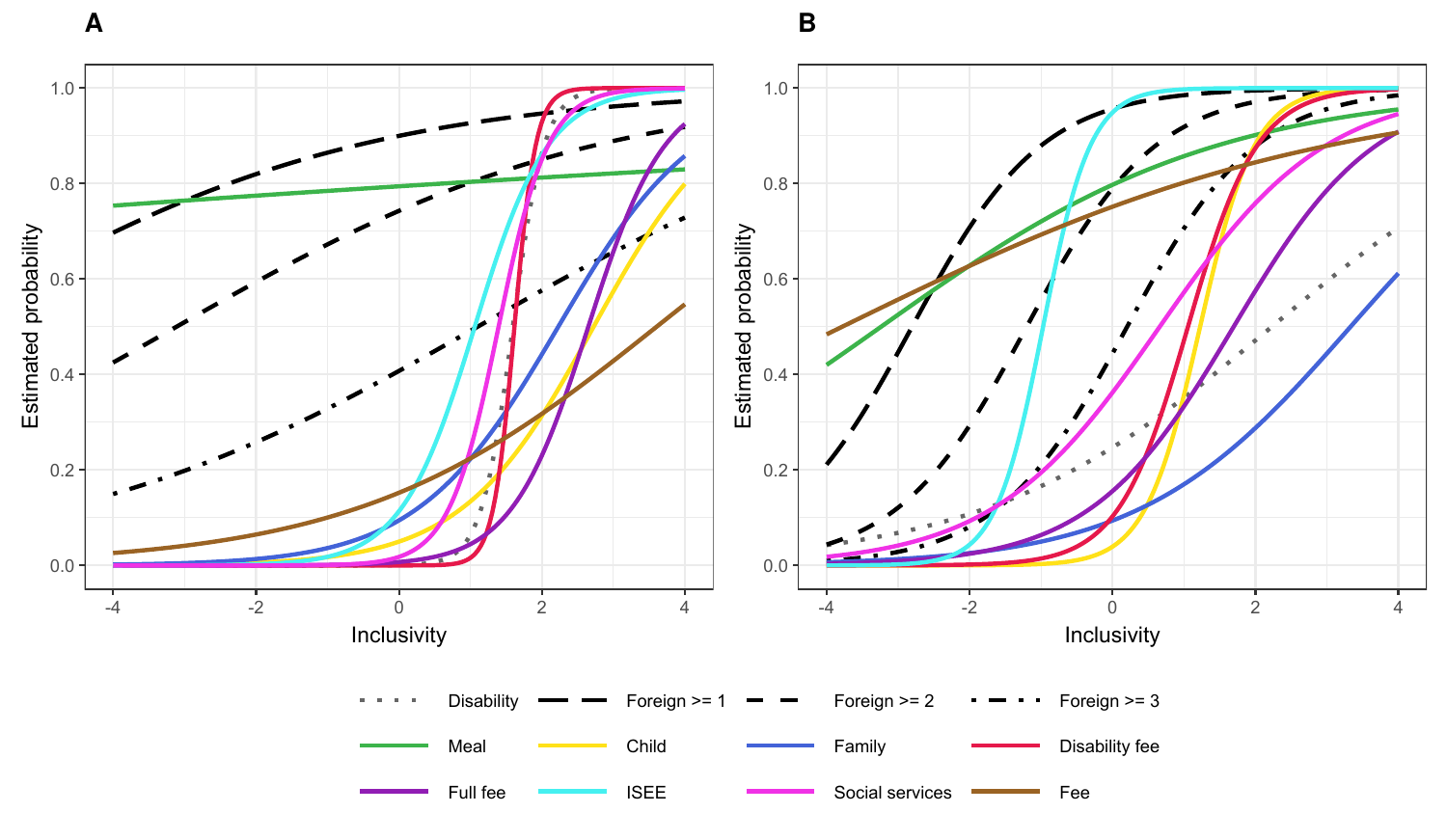}%
		\caption{Estimated item-response functions from Equations \eqref{model}-\eqref{model_foreign} for private (panel A) and public (panel B) ECEC services. For binary indicators, the curves represent the estimated probability of a positive response. For the ordinal \textit{Foreign} item, the three black curves represent the estimated cumulative probabilities of attaining category $r$ or higher, for $r \in \{1,2,3\}$.}\label{fig:prob_estimated}
	\end{figure}

	Figure~\ref{fig:prob_estimated} shows the estimated probabilities as functions of the corresponding latent dimension, separately for private (panel A) and public (panel B) services. For the binary indicators described in Table~\ref{tab:var}, the curves represent the probability of a positive response. For the ordinal \textit{Foreign} indicator, they represent the cumulative probabilities of being in category $r$ or higher, for $r\in\{1,2,3\}$. 
	
	The estimated item parameters and their standard errors for each ownership group are thus shown in Table~\ref{tab:estimated}. A constrained specification is adopted, whereby each item loads on a single latent dimension; the column ``Dim'' in Table~\ref{tab:estimated} indicates the corresponding dimension. To facilitate interpretation of the intercepts, Table~\ref{tab:estimated} also reports the estimated response probabilities when the corresponding latent dimension equals zero, i.e., its mean within each ownership group. For brevity, the column containing these probabilities is labeled $\hat{\pi}_{0}$.
	
	First of all, Figure~\ref{fig:prob_estimated} and Table~\ref{tab:estimated} show that all estimated discrimination parameters are positive, although their magnitudes vary. Thus, higher scores on each latent dimension are associated with higher probabilities of positive responses or, for \textit{Foreign}, of reaching higher categories. The direction of these associations is consistent with interpreting higher latent scores as greater inclusivity.
	
	Focusing on the social dimension, the estimates for private services require cautious interpretation. The \textit{Foreign} indicator has a modest discrimination estimate ($\hat\beta_1^F=0.341$). For \textit{Disability}, the estimated response probability at a latent score of zero is very low ($\hat\pi_0=0.001$), while the large discrimination estimate ($\hat\beta_1=4.524$) is accompanied by substantial uncertainty (standard error $3.611$). We nevertheless retain the social dimension because participation by children with disabilities and foreign children is an essential component of the definition of inclusivity adopted in this study.
	Excluding it would narrow the private-service index to financial provisions alone. Within public services, both discrimination parameters are estimated with lower relative uncertainty ($\hat\beta_1^F=1.102$, standard error $0.331$, for \textit{Foreign}; $\hat\beta_1=0.500$, standard error $0.167$, for \textit{Disability}), providing a clearer characterization of the social dimension.
	For \textit{Foreign}, the probability of reaching at least category 1 is already high at a latent score of zero ($0.956$), while the probabilities of reaching categories 2 and 3 are $0.790$ and $0.444$.
	Around the center of the dimension, this indicator therefore distinguishes mainly between higher levels of provincial foreign-child participation. Disability enrollment, instead, has an estimated probability of $0.247$ at zero and varies
	more gradually in log-odds along the dimension.
	
	Turning to the economic dimension, among private providers, reductions recommended by social services, ISEE-based reductions,
	and full-fee exemptions are strongly associated with higher
	economic inclusivity scores.
	Their estimated probabilities at a latent score of zero are
	$0.018$, $0.114$, and $0.007$, respectively, indicating that these provisions mainly characterize services further up the economic dimension. By contrast, meal inclusion in the fee has a high estimated probability at zero ($0.794$), but almost no association with the economic score ($\hat\beta_1=0.058$). It therefore provides little information about whether a private service also offers financial assistance to families facing economic disadvantage.
	Disability-related reductions are particularly discriminating, although the substantial uncertainty in their estimated discrimination parameter warrants caution. Within public services, ISEE-based reductions have both a high estimated probability at zero ($\hat\pi_0=0.949$) and a strong discrimination parameter ($\hat\beta_1=3.006$). Higher economic scores also identify services more likely to offer reductions related to children and disability: these provisions have low probabilities at zero ($\hat\pi_0=0.039$ and $0.103$, respectively), but their probabilities increase markedly along the economic dimension. Meal inclusion and the absence of a registration fee, although likely at zero, are less informative about these differences
	in financial support.
	
	Taken together, the results identify different priorities for strengthening inclusivity within the two ownership groups. Among public services, the high probability of ISEE-based reductions at the center of the economic dimension contrasts with the much lower probabilities of child- and disability-related reductions. This suggests that assistance based on household resources should be complemented by support addressing specific needs that income alone may not capture. Among private services, reductions recommended by social services, ISEE-based reductions, and full-fee exemptions characterize providers with higher economic inclusivity scores. Extending these provisions could therefore broaden the forms of financial protection available to disadvantaged families. The social results add a complementary set of policy objectives. Within private services, the uncertainty surrounding the estimates shows that social inclusivity remains less clearly captured. This calls for greater attention to its measurement and assessment, so that the evaluation of private provision adequately considers this important aspect alongside financial support. Within public services, the estimates provide a clearer picture. Nevertheless, the probability of enrolling a child with a disability is only $0.247$ for providers at the center of the social dimension. This identifies disability inclusion as an area warranting further assessment and support, taking account of local needs and demand. Finally, differences in provincial foreign-child participation are an important feature distinguishing social inclusivity within public services. These findings support giving explicit attention to both disability inclusion and foreign-child participation when assessing progress toward inclusive ECEC provision.
	
	\begin{table}[!h]
		
		\centering
		\resizebox{\ifdim\width>\linewidth\linewidth\else\width\fi}{!}{
			\fontsize{8}{10}\selectfont
			\begin{tabular}[t]{lccccccc}
				\toprule
				\multicolumn{2}{c}{ } & \multicolumn{3}{c}{Private} & \multicolumn{3}{c}{Public} \\
				\cmidrule(l{3pt}r{3pt}){3-5} \cmidrule(l{3pt}r{3pt}){6-8}
				Variable & Dim & $\hat{\beta}_0$ & $\hat{\beta}_1$ & $\hat{\pi}_{0}$ & $\hat{\beta}_0$ & $\hat{\beta}_1$ & $\hat{\pi}_{0}$\\
				\midrule
				Foreign ($Y\geq 1$) & 1 & 2.196 (0.119) & 0.341 (0.134) & 0.900 (0.011) & 3.089 (0.325) & 1.102 (0.331) & 0.956 (0.014)\\
				Foreign ($Y\geq 2$) & 1 & 1.061 (0.083) &  & 0.743 (0.016) & 1.322 (0.182) &  & 0.790 (0.030)\\
				Foreign ($Y\geq 3$) & 1 & -0.374 (0.072) &  & 0.408 (0.017) & -0.225 (0.115) &  & 0.444 (0.028)\\
				Disability & 1 & -7.203 (4.973) & 4.524 (3.611) & 0.001 (0.004) & -1.114 (0.115) & 0.500 (0.167) & 0.247 (0.021)\\
				\addlinespace
				Meal & 2 & 1.351 (0.086) & 0.058 (0.123) & 0.794 (0.014) & 1.370 (0.119) & 0.423 (0.143) & 0.797 (0.019)\\
				Fee & 2 & -1.717 (0.105) & 0.476 (0.135) & 0.152 (0.014) & 1.106 (0.108) & 0.293 (0.131) & 0.751 (0.020)\\
				Disability fee & 2 & -10.825 (4.836) & 6.702 (3.212) & 0.00002 (0.0001) & -2.168 (0.305) & 2.053 (0.402) & 0.103 (0.028)\\
				Full fee & 2 & -4.921 (0.526) & 1.858 (0.362) & 0.007 (0.004) & -1.691 (0.160) & 0.997 (0.194) & 0.156 (0.021)\\
				ISEE & 2 & -2.055 (0.207) & 1.939 (0.273) & 0.114 (0.021) & 2.920 (0.640) & 3.006 (0.835) & 0.949 (0.031)\\
				Child & 2 & -2.946 (0.207) & 1.082 (0.198) & 0.050 (0.010) & -3.196 (0.570) & 2.607 (0.629) & 0.039 (0.022)\\
				Social services & 2 & -4.000 (0.496) & 2.872 (0.461) & 0.018 (0.009) & -0.565 (0.111) & 0.856 (0.153) & 0.362 (0.026)\\
				Family & 2 & -2.261 (0.153) & 1.015 (0.166) & 0.094 (0.013) & -2.266 (0.181) & 0.680 (0.206) & 0.094 (0.015)\\
				\bottomrule
		\end{tabular}}
		\centering
		\caption{Estimated item parameters from the configural
			latent trait model. For binary indicators, $\hat{\beta}_0$ denotes
			the estimated intercept and $\hat{\beta}_1$ the estimated
			discrimination parameter. For the ordinal \textit{Foreign} item,
			$\hat{\beta}_0$ denotes the estimated cumulative intercept for each
			threshold; its discrimination parameter is common across the three
			cumulative logits and is reported once.
			The column $\hat{\pi}_0$ reports estimated probabilities evaluated
			at zero on the corresponding latent dimension, calculated as
			$\hat{\pi}_0=\operatorname{logit}^{-1}(\hat{\beta}_0)$.
			These are probabilities of a positive response for binary indicators
			and cumulative probabilities $\Pr(Y_{kg}\geq r\mid z_{1kg}=0)$
			for \textit{Foreign}, with $r\in\{1,2,3\}$.
			Standard errors are in parentheses.}\label{tab:estimated}
	\end{table}

	The estimated correlations between the social and economic
	dimensions were $0.400$ (standard error $0.094$) for private
	services and $0.834$ (standard error $0.176$) for public services.
	Therefore, a private provider's financial provisions offer
	an incomplete picture of its position on the social dimension. In contrast, 
	for public services, the stronger association suggests that
	the two dimensions more often reflect a shared pattern
	of inclusivity, although they describe different aspects
	of provision. Retaining the prespecified two-dimensional structure preserves
	this distinction between the conditions intended to support
	access and the participation captured by the social indicators,
	rather than treating the two as interchangeable.
	
\begin{figure}
		\centering
		\includegraphics[width=.4\columnwidth]{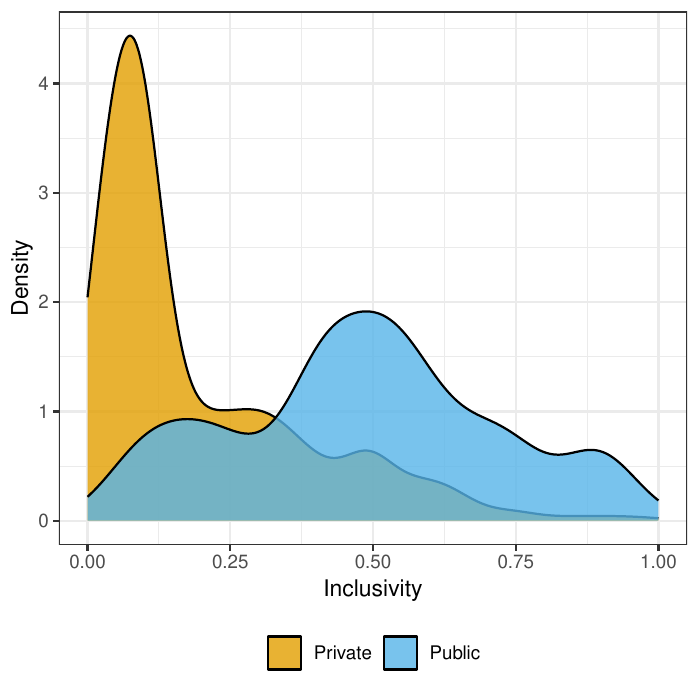}
		\caption{Density distributions of the normalized inclusivity index $\widetilde z$ defined in Equation \eqref{eq:scaledltm} for private (orange) and public (light blue) ECEC services. Scores are calibrated and rescaled separately within each ownership group; consequently, the horizontal positions and overlap of the two curves should not be interpreted as direct comparisons of absolute inclusivity levels.}
		\label{fig:hist}
	\end{figure}

	Figure~\ref{fig:hist} reveals different patterns of heterogeneity
	within the two ownership groups. Among private services, the pronounced right skew indicates that high inclusivity scores characterize a relatively small subset of providers, with most services concentrated at lower values. The approximately symmetric public-service distribution instead places intermediate scores at the center of the observed variation, with fewer providers at either extreme. Because the indices are calibrated and rescaled separately, the horizontal positions and overlap of the curves do not
	support comparisons of absolute inclusivity between ownership
	groups. These distributional patterns motivate examining whether
	territorial differences extend across the lower, central, and upper parts of each distribution, rather than focusing only on differences in the mean.
	
\begin{figure}
		\centering
		\includegraphics[width = .7\columnwidth]{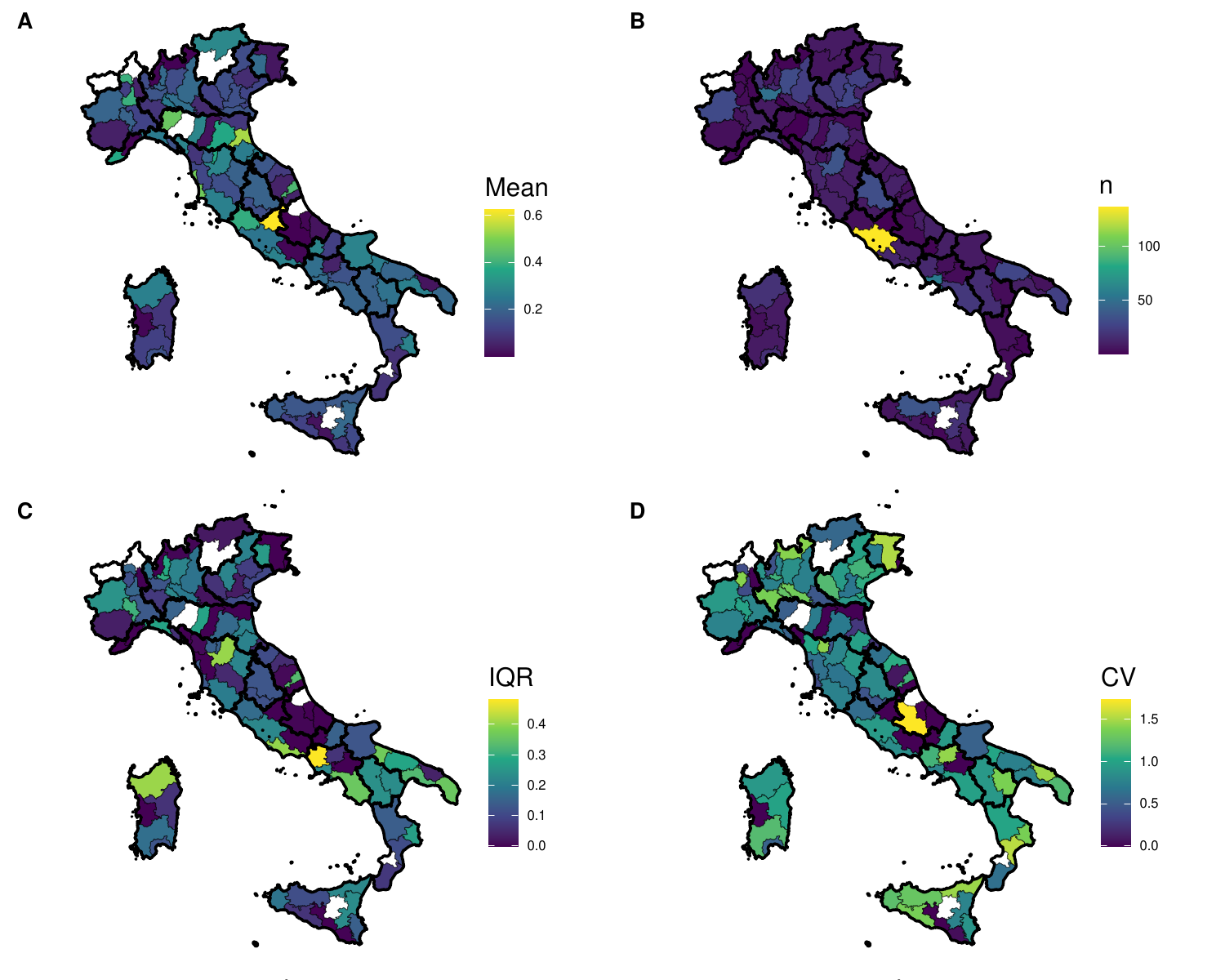}
		\caption{Mean (panel A), sample size $n$ (panel B), interquartile range (IQR; panel C), and coefficient of variation (CV; panel D) of the scaled inclusivity index $\tilde{z}$ (Equation~\eqref{eq:scaledltm}) considering private providers by Italian province. Black lines indicate regional boundaries. Darker shades correspond to lower values (closer to 0) and lighter shades to higher values; white denotes missing information.}
		\label{fig:median_private}
	\end{figure}

\begin{figure}
		\centering
		\includegraphics[width = .7\columnwidth]{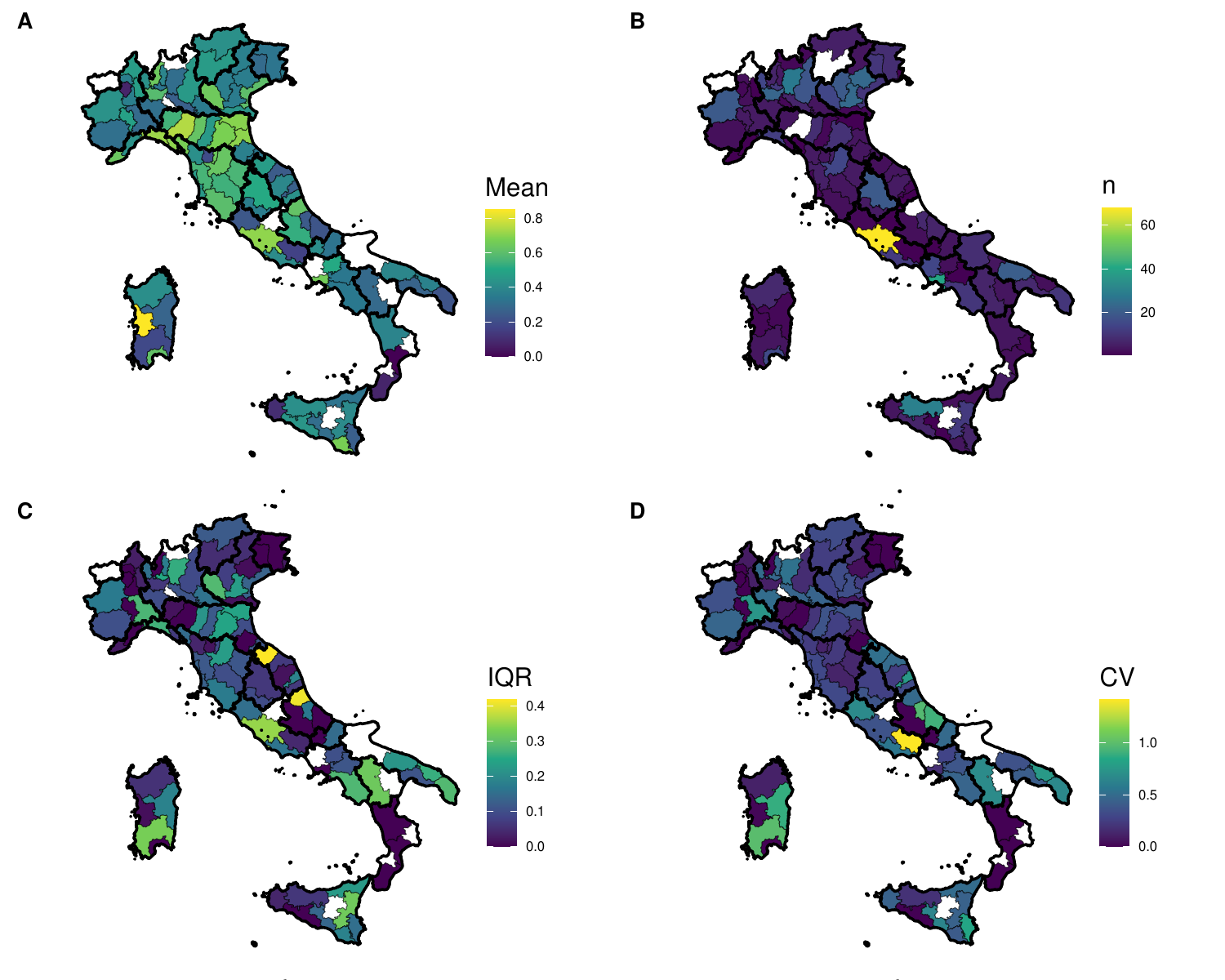}
		\caption{Mean (panel A), sample size $n$ (panel B), interquartile range (IQR; panel C), and coefficient of variation (CV; panel D) of the scaled inclusivity index $\tilde{z}$ (Equation~\eqref{eq:scaledltm}) considering public providers by Italian province. Black lines indicate regional boundaries. Darker shades correspond to lower values (closer to 0) and lighter shades to higher values; white denotes missing information.}
		\label{fig:median_public}
	\end{figure}

	Instead, Figures~\ref{fig:median_private}--\ref{fig:median_public} display the mean, sample size ($n$), interquartile range (IQR), and absolute coefficient of variation (CV) values of $\tilde{z}$ at the province level considering the private and public providers, respectively, where the black segments represent the regional boundaries. We can observe strong between- and within-region variability; e.g., the Lazio region has provinces with low levels of inclusivity as well as high ones. 
	
	Figures~\ref{fig:median_private}-\ref{fig:median_public} require cautious interpretation. Provincial means of $\tilde{z}$  may be unreliable in the presence of small or highly unbalanced sample sizes. As shown in the top-right panel of Figures~\ref{fig:median_private}-\ref{fig:median_public}, the number of sampled services per province ranges from $1$ (e.g., in some provinces of Emilia-Romagna and Lazio) to $136$ (Rome), indicating substantial imbalance across domains. The bottom panels further illustrate dispersion through the IQR and the CV. Both very low and very high CV values are observed within the same region (e.g., Friuli Venezia Giulia), while Sardinia records the highest IQR. Although several provinces display low CV, their interpretation should remain cautious, as the CV is mechanically related to the magnitude of the corresponding point estimate and may be unstable in small samples.
	
\begin{figure}
		\centering
		\includegraphics[width=.8\columnwidth]{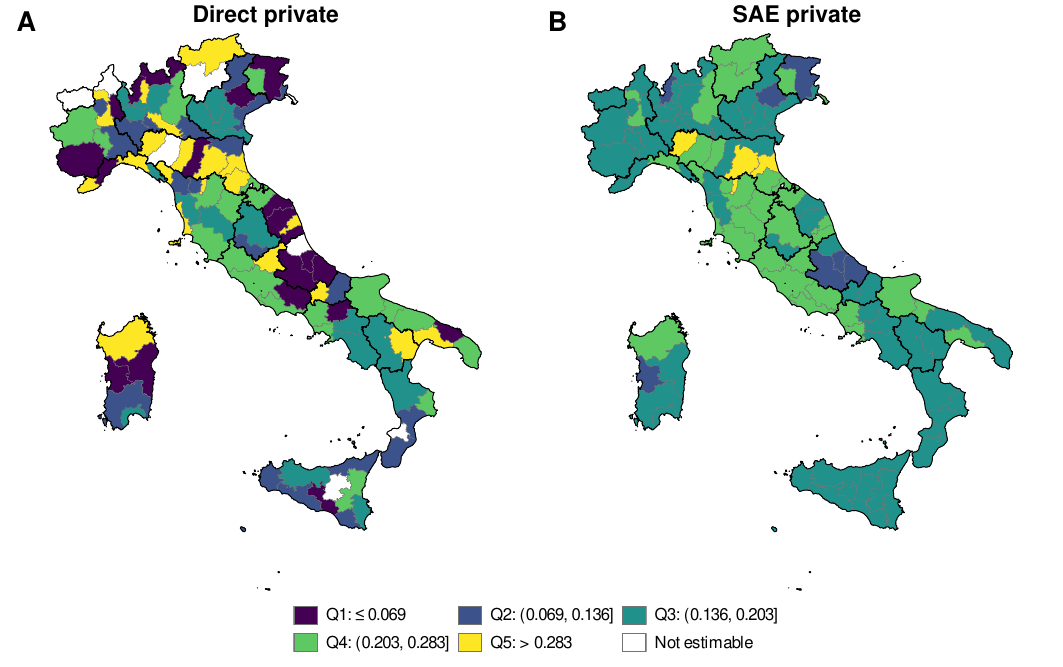}
		\caption{Provincial direct estimates (panel A) and census SAE EBPs (panel B) of the normalized inclusivity index $\tilde{z}$ (Equation~\eqref{eq:scaledltm}) for private ECEC services. Direct estimates are available for 99 provinces, whereas census SAE EBPs are obtained for all 107 provinces. Both panels use the same five classes, defined by the quintiles of the distribution of the 99 available provincial direct estimates. Darker shades indicate lower values and lighter shades higher values. Black lines denote regional boundaries; white areas indicate provinces without a direct estimate.}
		\label{fig:ebp_priv}
	\end{figure}

\begin{figure}
		\centering
		\includegraphics[width=.8\columnwidth]{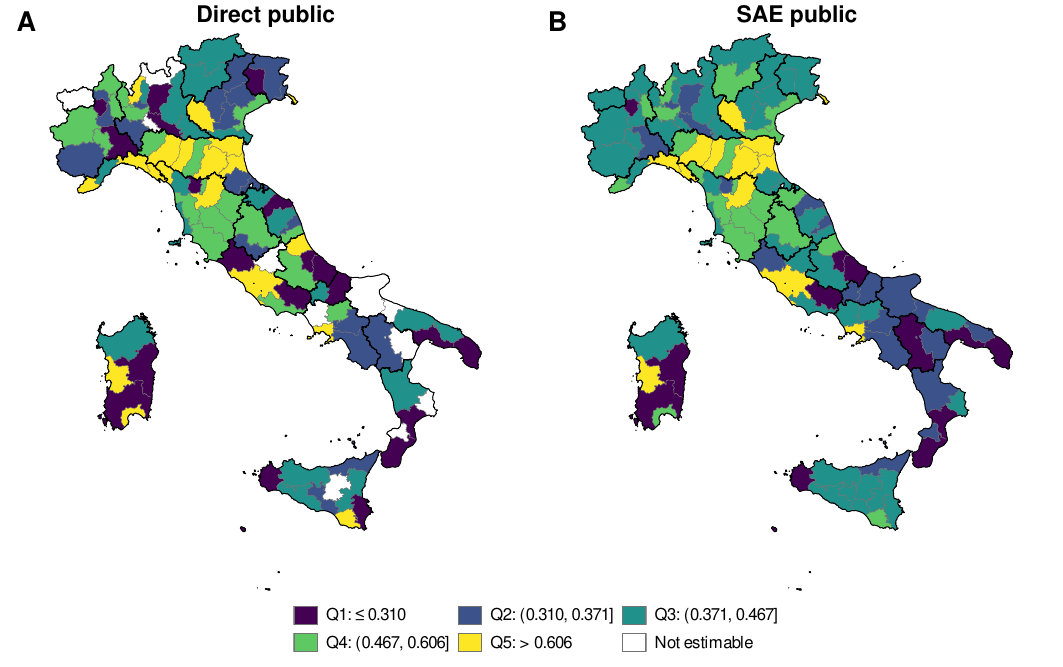}
		\caption{Provincial direct estimates (panel A) and census SAE EBPs (panel B) of the normalized inclusivity index $\tilde{z}$ (Equation~\eqref{eq:scaledltm}) for public ECEC services. Direct estimates are available for 96 provinces, whereas census SAE EBPs are obtained for all 107 provinces. Both panels use the same five classes, defined by the quintiles of the distribution of the 96 available provincial direct estimates. Darker shades indicate lower values and lighter shades higher values. Black lines denote regional boundaries; white areas indicate provinces without a direct estimate.}
		\label{fig:ebp_pub}
	\end{figure}

	To address these limitations at the provincial level, Figures~\ref{fig:ebp_priv} and~\ref{fig:ebp_pub} compare direct estimates (panels A) with census SAE EBP (panels B) for private and public ECEC services, respectively. Both panels within each figure use the same five classes, defined by the quintiles of the distribution of the available provincial direct estimates. These thresholds are calculated separately for private and public services, giving each province equal weight. Consequently, colors allow direct and model-based estimates to be compared only within each ownership group. Direct estimates are available for 99 private-service and 96 public-service provinces, whereas the SAE approach provides predictions for all 107 provinces in each group. Relative to the direct estimates, the model-based maps display a less fragmented spatial pattern, consistent with the borrowing-of-strength mechanism of the BHF model in Equation~\eqref{eq:bhf}. This concentration is particularly evident among private services: 96 of the 107 census SAE EBPs fall within the third and fourth classes defined by the direct-estimate quintiles. Because the class boundaries are held fixed, this pattern reflects the narrower distribution of the predictions rather than a change in the mapping scale. The public-service predictions remain distributed across all five classes. Some descriptive diagnostics comparing census SAE EBPs with direct estimates are reported in Appendix~\ref{app:diagnosticsae-inclusivity}.
	
	Table~\ref{tab:bootstrap_rrmse} complements Figures~\ref{fig:ebp_priv} and~\ref{fig:ebp_pub} by comparing their relative uncertainty. Comparisons are restricted to the same 99 private-service and 95 public-service provinces with both measures available; the public-service direct estimate for the Catanzaro province is zero, making its CV undefined. Median CV and Relative Root Mean Squared Error (RRMSE) decrease from 44.45\% for direct estimates to 31.59\% for census SAE EBPs among private services, and from 21.47\% to 18.13\% among public services. The upper quartiles and maxima are also lower for census SAE EBPs. Using the Statistics Canada thresholds as a descriptive classification \citep{canada}, the proportion exceeding 33.3\% decreases from 58.6\% to 44.4\% among private services and from 23.2\% to 4.2\% among public services.

	\begin{table}[!t]
		\centering
		\small
		\caption{Distribution of relative uncertainty in provincial direct estimates and census SAE EBPs of the inclusivity index, separately for private and public ECEC services. Comparisons are restricted to provinces with both measures available. Panel A reports the minimum, quartiles, and maximum; Panel B reports the number and percentage of provinces in each variability class.  For direct estimates, $\mathrm{CV}=100\sqrt{\widehat{\mathrm{Var}}
				(\widehat\mu^{\mathrm{dir}})}/|\widehat\mu^{\mathrm{dir}}|$;
			for census SAE EBPs, $\mathrm{RRMSE}=100\sqrt{\widehat{\mathrm{MSE}}_{\mathrm{boot}}
				(\widehat\mu^{\mathrm{EBP}})}/|\widehat\mu^{\mathrm{EBP}}|$,
			where $\widehat\mu^{\mathrm{dir}}$ and $\widehat\mu^{\mathrm{EBP}}$ denote the direct estimate and census SAE EBP, respectively, and $\widehat{\mathrm{MSE}}_{\mathrm{boot}}$ denotes the bootstrap estimate of mean squared prediction error. Thresholds follow Statistics Canada \citep{canada} and are applied to RRMSE as a descriptive convention. Quantiles are unweighted across provinces. Percentages in Panel B use the column-specific number of provinces.}
		\label{tab:bootstrap_rrmse}
		
		\begin{tabular*}{.95\columnwidth}
			{@{\extracolsep{\fill}}lrrrr@{}}
			\toprule
			& \multicolumn{2}{c}{Private}
			& \multicolumn{2}{c}{Public} \\
			\cmidrule(lr){2-3}
			\cmidrule(lr){4-5}
			& Direct CV & SAE EBP RRMSE
			& Direct CV & SAE EBP RRMSE \\
			\midrule
			
			\multicolumn{5}{l}{
				\textit{Panel A. Distribution of relative uncertainty (\%)}} \\
			Minimum        & 8.53    & 8.40  & 3.58   & 2.97  \\
			First quartile & 24.47   & 21.68 & 15.24  & 13.46 \\
			Median         & 44.45   & 31.59 & 21.47  & 18.13 \\
			Third quartile & 105.11  & 41.44 & 32.47  & 24.66 \\
			Maximum        & 6130.23 & 65.17 & 215.01 & 46.56 \\
			
			\midrule
			\multicolumn{5}{l}{
				\textit{Panel B. Variability classes: number (\%) of provinces}} \\
			$\leq 16.6\%$
			& 8 (8.1)   & 6 (6.1)   & 32 (33.7) & 37 (38.9) \\
			$16.6\% < x \leq 33.3\%$
			& 33 (33.3) & 49 (49.5) & 41 (43.2) & 54 (56.8) \\
			$>33.3\%$
			& 58 (58.6) & 44 (44.4) & 22 (23.2) & 4 (4.2)   \\
			\bottomrule
		\end{tabular*}
		\par\smallskip
		\begin{minipage}{.95\columnwidth}
		\end{minipage}
	\end{table}
	
	To sum up, relative uncertainty is generally lower for census SAE EBPs than for direct estimates across the reported summaries, although
	appreciable uncertainty remains, particularly among private services. Figures~\ref{fig:ebp_priv} and~\ref{fig:ebp_pub} also reveal substantial territorial variation within each ownership group. Among private services, relatively high predictions occur in several provinces of Emilia-Romagna and Tuscany, whereas lower predictions are also found in Northern provinces, including Como, Treviso, and Udine. Among public services, relatively high predictions occur in provinces such as Ravenna, Genova, and Roma, but also in Oristano and Napoli. Conversely, several Southern and insular provinces occupy lower classes. These patterns indicate that broad North--South differences coexist with considerable variation within macro-areas and regions, rather than defining a uniform geographical divide.
	
	Given the index distribution and the strong variability within regions shown in Figures \ref{fig:hist}--\ref{fig:ebp_pub}, the following subsection presents the results of the quantile model, including regional-level random effects.
	
	\subsection{Regional patterns in conditional inclusivity}\label{virt}
	
	The random-effects $\tau$-quantile model with $\tau \in \{0.25, 0.5, 0.75\}$ is estimated separately by ECEC service ownership (i.e., public/private) with the regions as a random effect.
	
\begin{figure}
		\centering
		\includegraphics[width=\columnwidth]{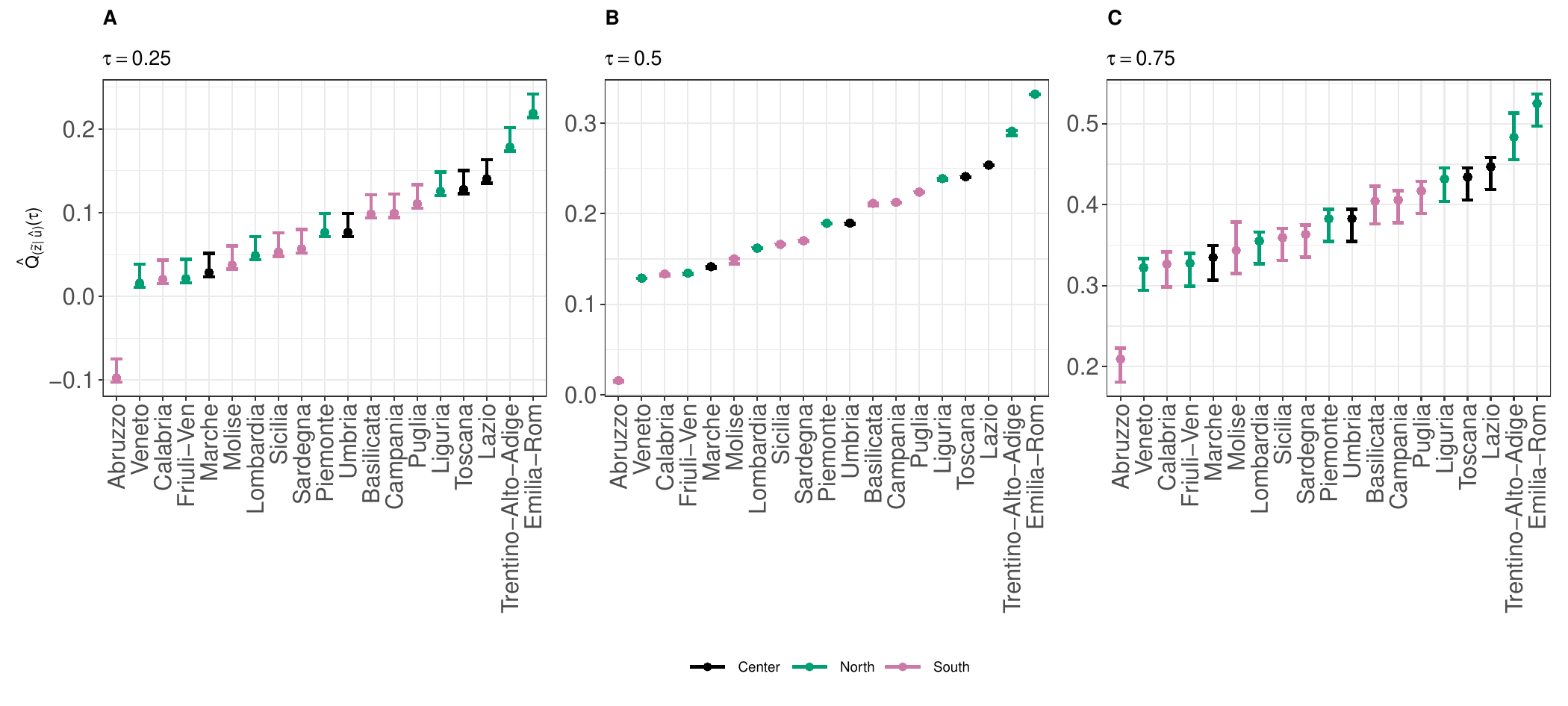}
		\caption{Conditional quantile  estimates based on Equation \eqref{eq:quantile_model} for private ECEC services at level $\tau \in \{0.25, 0.5, 0.75\}$. Points denote estimated conditional quantiles and vertical bars denote 95\% bootstrap confidence intervals. Colors identify the macro-area to which each region belongs.}
		\label{fig:quantile_res_private}
	\end{figure}

\begin{figure}
		\centering
		\includegraphics[width=\columnwidth]{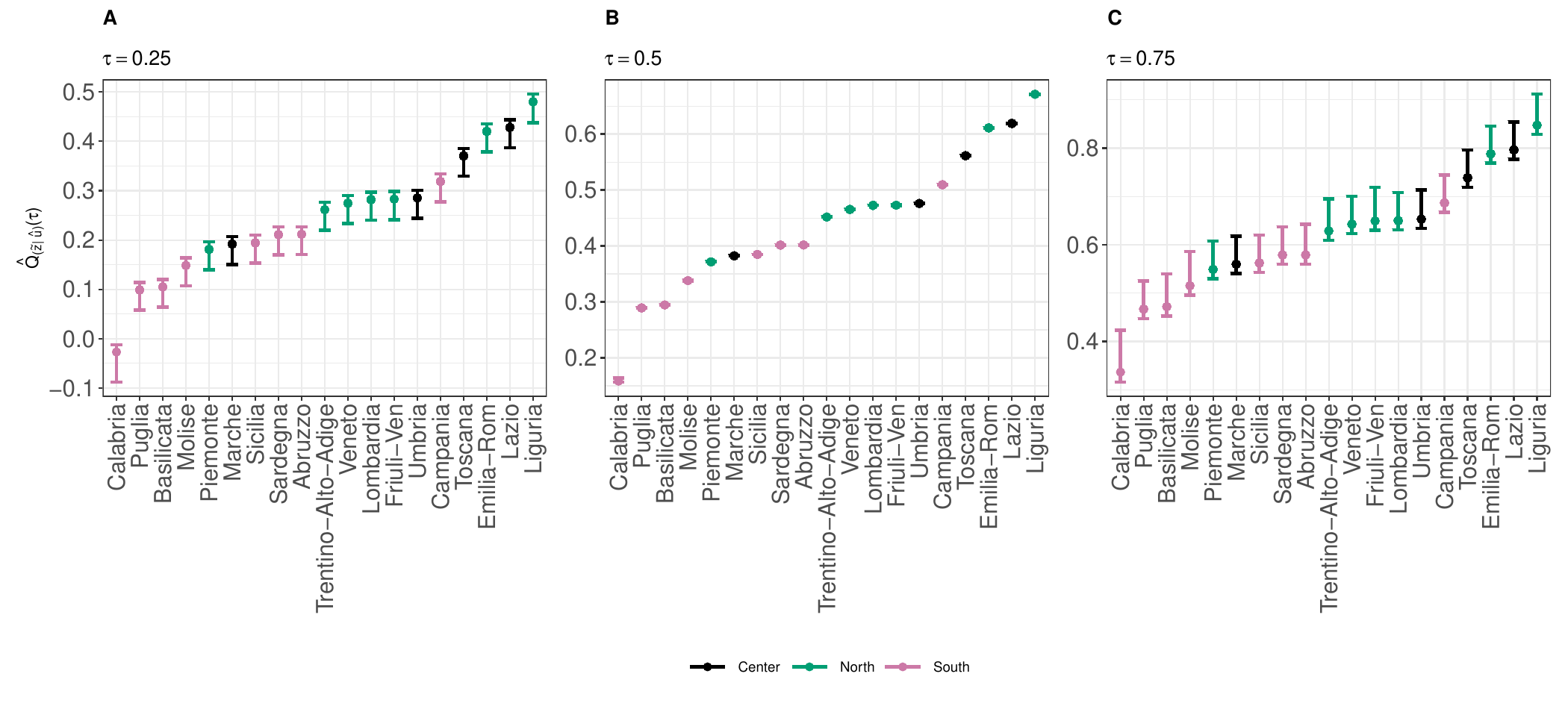}
		\caption{Conditional quantile  estimates based on Equation \eqref{eq:quantile_model} for public ECEC services at level $\tau \in \{0.25, 0.5, 0.75\}$. Points denote estimated conditional quantiles and vertical bars denote 95\% bootstrap confidence intervals. Colors identify the macro-area to which each region belongs.}
		\label{fig:quantile_res_public}
	\end{figure}

	Figures~\ref{fig:quantile_res_private}--\ref{fig:quantile_res_public} report the region-specific conditional quantile estimates at $\tau\in\{0.25,0.50,0.75\}$, respectively for private and public ECEC services, together with 95\% bootstrap confidence intervals. Uncertainty is assessed through a block bootstrap with $R=100$ resamples obtained by sampling regions with replacement, thereby preserving within-region dependence \citep{geraci2007quantile}. Among private services, Emilia-Romagna remains at the upper end of the ownership-specific distribution at all three quantiles, followed by Trentino--Alto Adige, while Lazio, Tuscany, and Liguria also occupy relatively high positions. Abruzzo is clearly separated at the lower end, with Veneto, Calabria, and Friuli--Venezia Giulia also located in the lower part of the distribution. Among public services, Liguria consistently occupies the upper end, followed by Lazio and Emilia-Romagna, while Tuscany also shows relatively high estimates. Calabria is located at the lower end, followed by Puglia and Basilicata. Because the linear predictor in Equation \eqref{eq:quantile_model} is not constrained to the unit interval, fitted conditional quantiles at the lower tail may fall below zero for regions located near the boundary of the observed distribution. Such estimates are interpreted as boundary extrapolations indicating a very low relative position within the ownership-specific distribution, rather than as substantively negative levels of inclusivity.
	
	The following discussion provides an institutional interpretation of the regional patterns rather than a causal explanation identified by the fitted models.
	
	In the case of Lazio, the prominent role of the municipality of Rome provides a plausible explanation for the estimated regional pattern. Nearly half of the region's children aged 0--2 reside in the capital, placing a substantial share of the target population under a single regulatory and administrative framework. This centralized governance structure oversees public, private, and affiliated providers, promoting greater coordination and standardization in service provision. The concentration of both the target population and service governance in the metropolitan area offers an institutional explanation for Lazio's stable upper position within both ownership-specific distributions.
	
	Liguria's high relative position is consistent with its long-standing investment in social policies and its strong tradition of community-based service provision. The region is characterized by a diversified network of providers, including public institutions and private non-profit organizations such as social cooperatives. These actors often operate with explicit social inclusion objectives, which can enhance accessibility and territorial coverage. The interaction between institutional commitment and a pluralistic provider structure creates an institutional setting favorable to accessibility and the adoption of inclusion-related measures.
	
	Abruzzo, by contrast, occupies the lowest position in the private-service distribution. The limited availability of publicly provided ECEC services increases reliance on private alternatives, which may be unevenly distributed and financially inaccessible to a substantial share of households. These challenges are compounded by fiscal limitations and administrative capacity constraints, which can restrict the effective absorption and implementation of national and EU funds intended to expand childcare infrastructure. Together, these structural and institutional factors contribute to sustained disparities in access to quality ECEC services within the region.
	
	The observed regional variation in the inclusivity and configuration of ECEC services may reflect heterogeneity in local policy frameworks and the role played by regional governments in regulating and financing the system. In particular, part of these disparities may be related to variations in per capita public expenditure allocated to ECEC, which are likely to contribute to uneven territorial development in terms of coverage and accessibility. In some regions, the organization of service provision may reflect historical trajectories characterized by the prominent presence of ecclesiastical and Christian-inspired lay organizations. This institutional legacy may have influenced the current structure of the supply system, often embedding ECEC services within pre-existing social and community networks and shaping patterns of provision over time. Another relevant factor concerns the demand for ECEC services, which is strongly associated with female labor force participation rates and varies substantially across regions. Higher employment rates among women are typically associated with greater demand for ECEC services and with the expansion and configuration of local supply (see \citet{andreella2024spatial} for a more detailed discussion). More broadly, in the absence of clear and binding national guidelines and, in some cases, well-defined regional coordination, the development of the ECEC system appears to have evolved largely in response to local demand conditions. As a consequence, ECEC services may have increasingly assumed a work--life balance orientation, often being framed more as social support mechanisms than as components of a universal educational pathway. This interpretation is consistent with municipal admission regulations, which frequently prioritize criteria related to parental employment status.

	\section{Discussion}\label{discussion}
	
	This study conceptualizes inclusivity in Italian ECEC services as a multidimensional construct encompassing both social openness (e.g., enrollment of foreign children and children with disabilities) and economic accessibility (e.g., fee reductions and exemptions). The latent trait model defines this construct and shows different inclusivity patterns in public and private services. In particular, the model suggests complementing income-based assistance with support for specific child- and disability-related needs in public services, and extending the availability of targeted reductions and exemptions in private services. Recognizing these differences is particularly relevant for policy-makers given the limited availability of public provision and the growing role of private care in Italy \citep{da2013nothing}. In addition, the quantile regression analysis shows substantial territorial heterogeneity, with several northern and central regions occupying higher positions and some southern regions at the lower end. Regional rankings remain broadly stable across quantiles, i.e., territorial differences concern services with low, intermediate, and high inclusivity scores. However, these rankings differ between public and private services, highlighting again the importance of analyzing the two sectors separately. Finally, the census SAE EBP approach complements this analysis by providing model-based estimates of ownership-specific provincial mean inclusivity, thereby extending inference to domains not explicitly targeted in the sampling design. The resulting indicator describes a the relative inclusivity within each ownership group at NUTS-3 level while explicitly quantifying estimation uncertainty.

	Policymakers should consider these findings when designing interventions to enhance inclusivity, ensuring that both public and private providers contribute to equitable access. To effectively address the regional disparities highlighted in our analysis, a nationally coordinated approach is essential. We argue that targeted policies aimed at enhancing the participation of the most vulnerable population groups, such as direct subsidies for low-income families or the establishment of mandatory inclusivity criteria for all providers, could be defined and implemented through the ``Intesa Stato-Regioni'' (State-Regions Agreement). This approach is crucial as it represents the appropriate institutional mechanism within Italy's quasi-federal system to harmonize social policies, ensuring that efforts to improve equity are consistently applied across all regions while respecting their administrative autonomy.
	
	The proposed study has some limitations. First, the analysis follows a sequential approach. This structure makes the construction of the index and its subsequent territorial analysis transparent, and allows each component to address a distinct research objective. However, the associated uncertainty is not considered except for the SAE analysis. The literature also supports this two-step approach. Studies such as those by \citet{skrondal2001regression, kaplan2008structural} discuss conditions under which regressions on factor scores can provide useful approximations, with measurement error absorbed into the regression disturbance. Nonetheless, extending these results to distributional models (such as quantile regression) may require additional assumptions, and future work could integrate measurement and regression in a unified framework.

	Second, inclusivity is inherently broader than the dimensions observable in the available survey data. The analysis is restricted to social and economic indicators captured in the Italian ECEC survey. Other information, such as
	the presence of qualified staff \citep{devore2000wanted}, ECEC service hours \citep{stahl2020early}, or more detailed information on meal provision (beyond whether meals are included
	in the fee), may further refine the measurement of inclusivity. For instance, staff qualifications are crucial for ensuring equitable learning opportunities, while flexible service hours and meal availability can address barriers faced by working parents or families with specific needs. Information on barriers experienced by families, including those who do not use ECEC services, would also help distinguish the availability of inclusion-related provisions from their effectiveness in enabling access. However, since our analysis relies on the Italian ECEC survey, the measurement is necessarily limited to the indicators it collects. Incorporating the additional dimensions mentioned above would allow a more comprehensive assessment of inclusivity in ECEC services. Nevertheless, the social and economic dimensions of inclusivity explored in this study offer a starting point for analyzing the Italian case beyond what coverage indicators alone can reveal. Further research is needed to delve deeper into the complexities of other factors influencing inclusivity. Societal, philosophical, and economic debates surrounding the broader concept of inclusivity warrant exploration in future studies, contributing to a holistic understanding of ECEC services if additional data becomes available.

	\section*{Declaration of Competing Interest}
	The authors declare no competing interests.
	
	\section*{Acknowledgments}
	The authors acknowledge the agreement signed among the University Ca' Foscari of Venice, the National Institute of Statistics, and the Department of Family Policies at the Presidency of the Ministerial Council that supported and partially funded their work.
	
	\section*{Data availability}
	The data supporting the findings of this study are available from the corresponding author upon reasonable request.
	
	\section*{Funding statement}
	This research received no external funding.
	
	\section*{Author Contributions}
	\textbf{AA}: conceptualization, data curation, formal analysis, investigation, writing - original draft, and writing - review \& editing., \textbf{GB}: conceptualization, formal analysis, writing - review \& editing, \textbf{FC}: conceptualization, writing - review \& editing, \textbf{SC}: conceptualization, writing - review \& editing and supervision.
	
	\clearpage
	\appendix

		\section{Exploratory data analysis}\label{app:descriptive}
		
\begin{figure}
			\centering
			\includegraphics[width=.9\columnwidth]{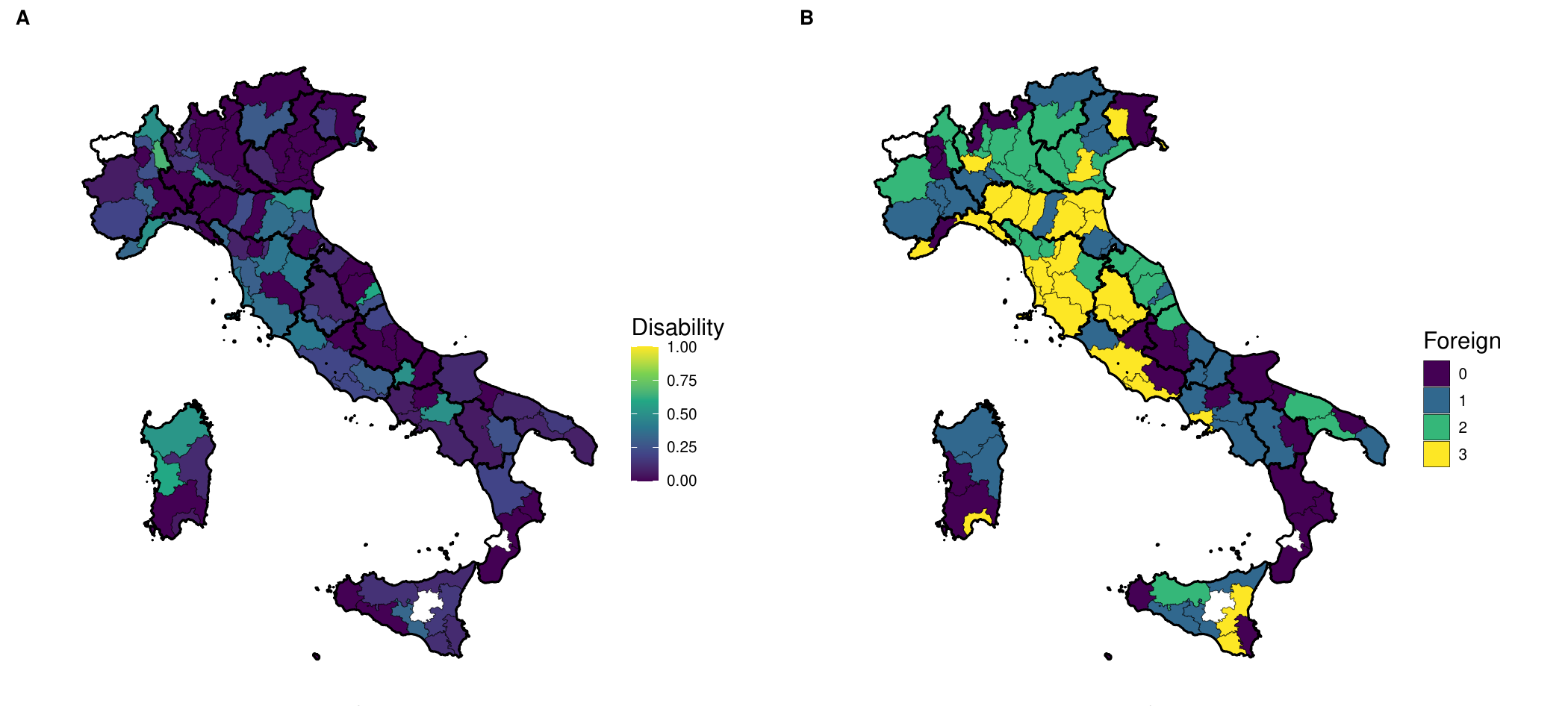}
			\caption{Provincial distribution of the \textit{Disability} (panel A) and \textit{Foreign} (panel B) indicators. Black lines denote regional boundaries. Darker shades indicate a value of 0, lighter shades a value of 1 for \textit{Disability} and 3 for \textit{Foreign}, and white denotes missing information.}
			\label{fig:DIS_FOR}
		\end{figure}

		An essential aspect of defining an ECEC service as inclusive is its ability to reach all children \citep{bulgarelli2019nido}. Italian Law 104/92 emphasizes the right to education from birth, granting priority access to children aged 0 to 3 with disabilities. Municipalities are mandated to establish inclusive services, and this obligation may extend to private ECEC providers through formal agreements. In the 2021/2022 academic year, \citetalias{Istat2022} reports that $13.5\%$ of ECEC supply units in Italy admitted children with certified disabilities. However, substantial variation persists between public and private services and across Italian regions. Families with children with disabilities may often prefer public services not only for financial reasons but also because these institutions are explicitly required to guarantee access, frequently offering free enrollment and targeted support measures. The macro-area of Central Italy shows a comparatively higher propensity among private providers to admit children with disabilities ($14.2\%$) than in the North and South. Panel A of Figure~\ref{fig:DIS_FOR} illustrates the provincial distribution of the \textit{Disability} variable defined in Table~\ref{tab:var}.
		
		Another critical dimension of inclusivity concerns the enrollment of foreign children. The presence of children without Italian citizenship in ECEC has increased over the past two decades. In 2002, approximately $4\%$ of children up to age six did not hold Italian citizenship; this share rose steadily until 2013 and has since stabilized at around $14\%$ of the total resident population in that age group. This national figure conceals substantial territorial disparities, with foreign children more concentrated in the North (around $20\%$), followed by the Center (approximately $16\%$), and a considerably lower presence in the South (around $6\%$). In the 2021/2022 academic year, foreign children represented about $6.7\%$ of ECEC attendees, with the highest shares observed in public services in Northern regions ($14\%$) (ISTAT et al., \citeyear{Istat2022}). \citetalias{Istat2022} highlights the existence of barriers limiting access for foreign families, particularly cost-related constraints. Access to formal ECEC services plays a crucial role in the social integration of immigrant families and in reducing inequalities in access to social services \citep{bonizzoni2014}. Panel B of Figure\ref{fig:DIS_FOR} displays the provincial distribution of the \textit{Foreign} variable defined in Table~\ref{tab:var}.
		
\begin{figure}
			\centering
			\includegraphics[width=\columnwidth]{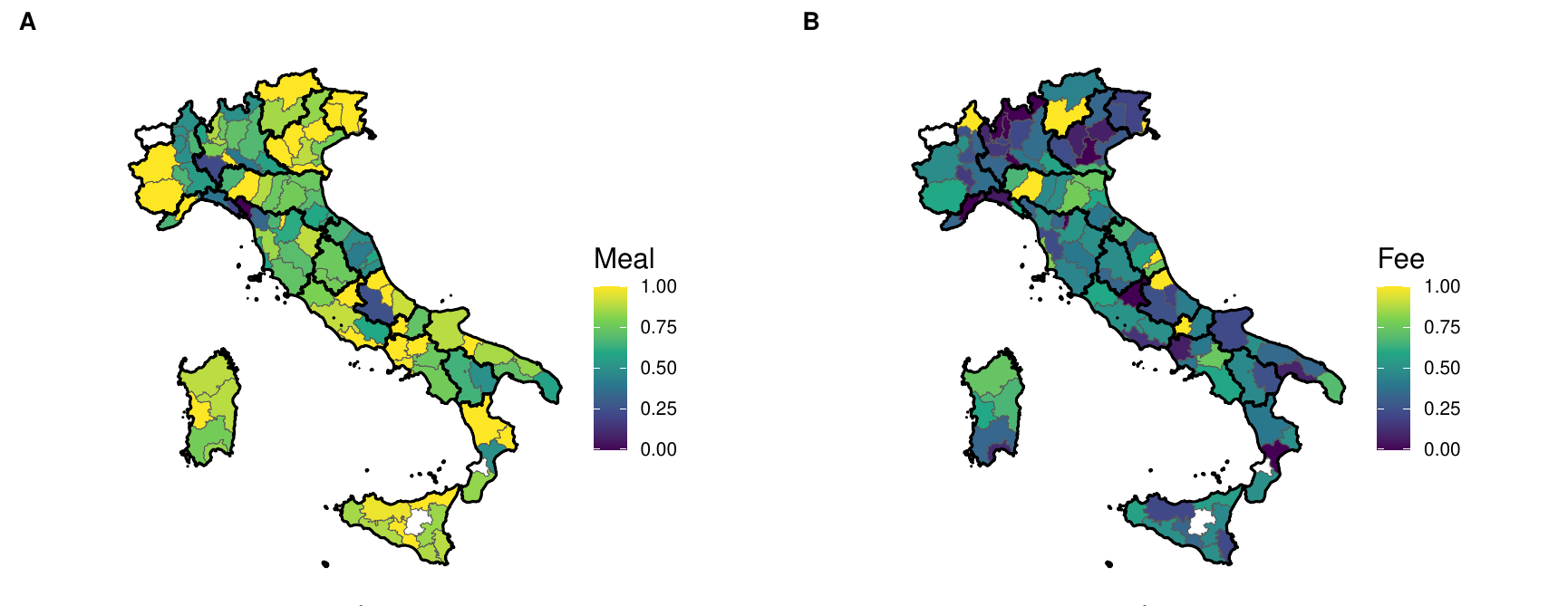}
			\caption{Provincial distribution of the \textit{Meal} (panel A) and \textit{Fee} (panel B) indicators. Black lines denote regional boundaries. Darker shades indicate a value of 0, lighter shades a value of 1, and white denotes missing information.}
			\label{fig:PASTO}
		\end{figure}

\begin{figure}
			\centering
			\includegraphics[width=.8\columnwidth]{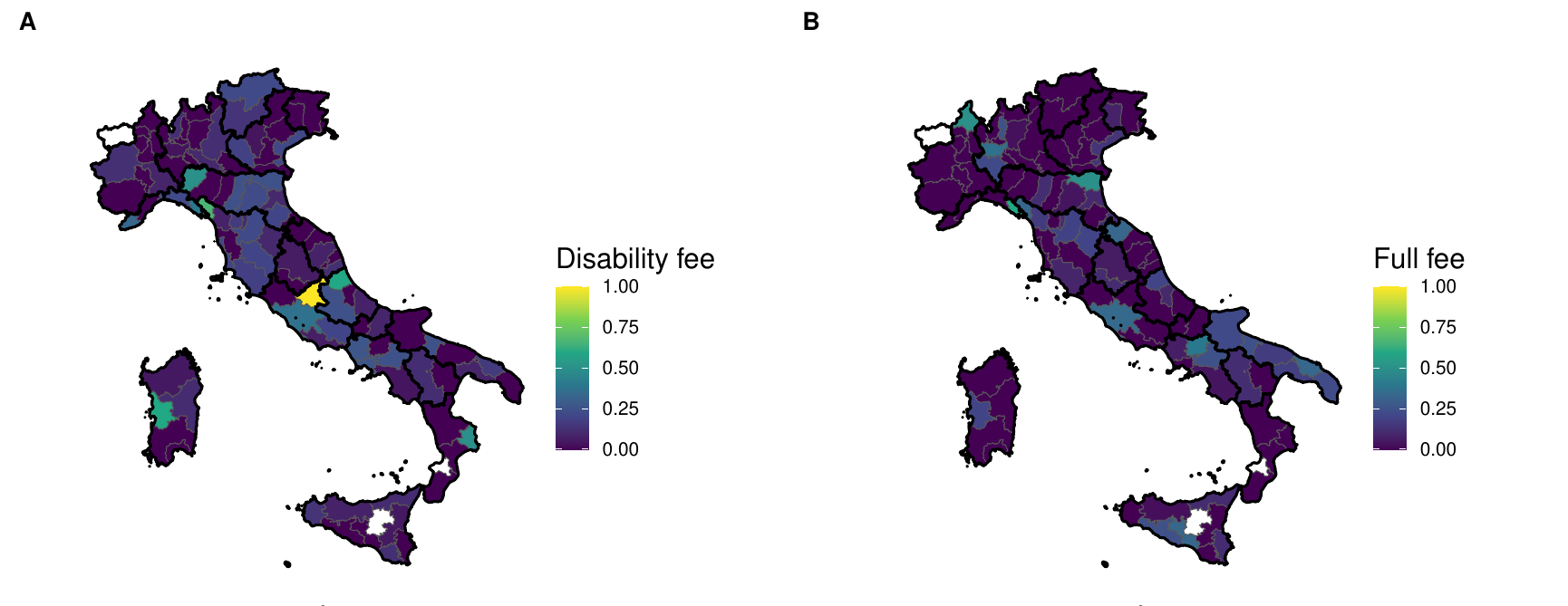}
			\caption{Provincial distribution of the \textit{Disability fee} (panel A) and \textit{Full fee} (panel B) indicators. Black lines denote regional boundaries. Darker shades indicate a value of 0, lighter shades a value of 1, and white denotes missing information.}
			\label{fig:RETTE_DISABILITA_PREVISTO}
		\end{figure}

\begin{figure}
			\centering
			\includegraphics[width=\columnwidth]{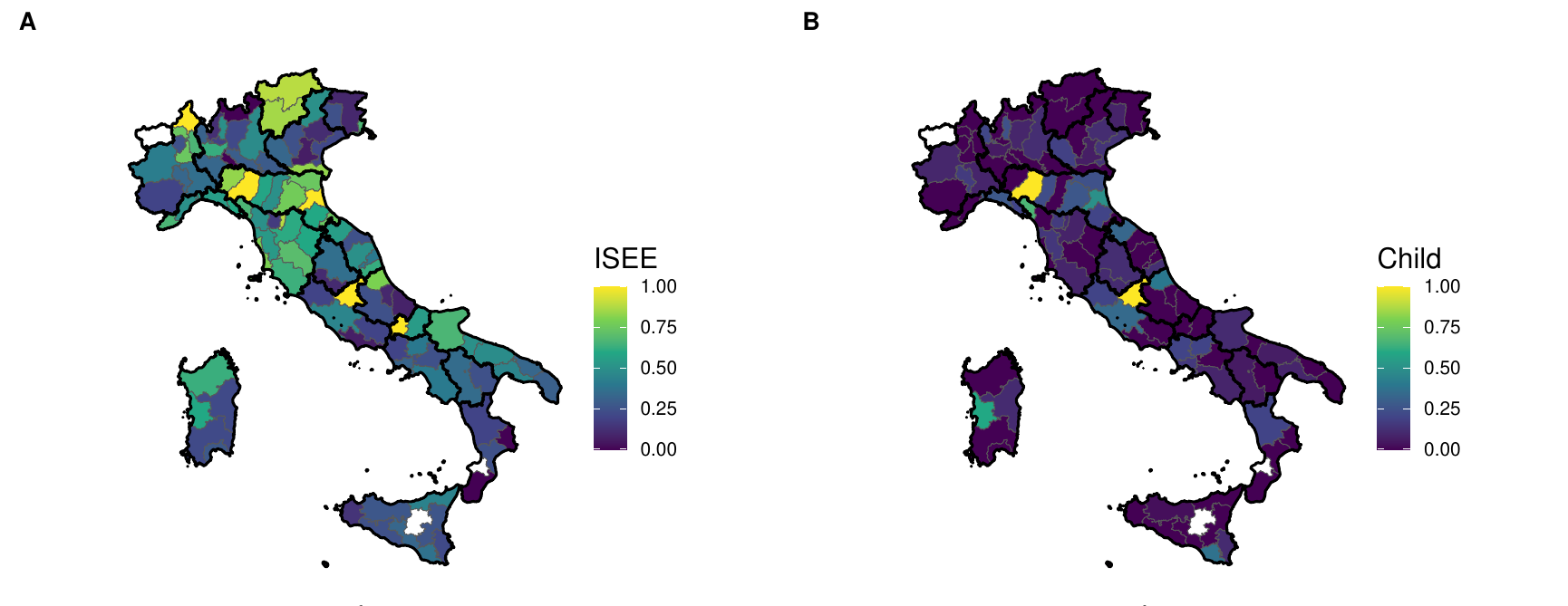}
			\caption{Provincial distribution of the \textit{ISEE} (panel A) and \textit{Child} (panel B) indicators. Black lines denote regional boundaries. Darker shades indicate a value of 0, lighter shades a value of 1, and white denotes missing information.}
			\label{fig:RETTE_ISEE_PREVISTO}
		\end{figure}

\begin{figure}
			\centering
			\includegraphics[width=\columnwidth]{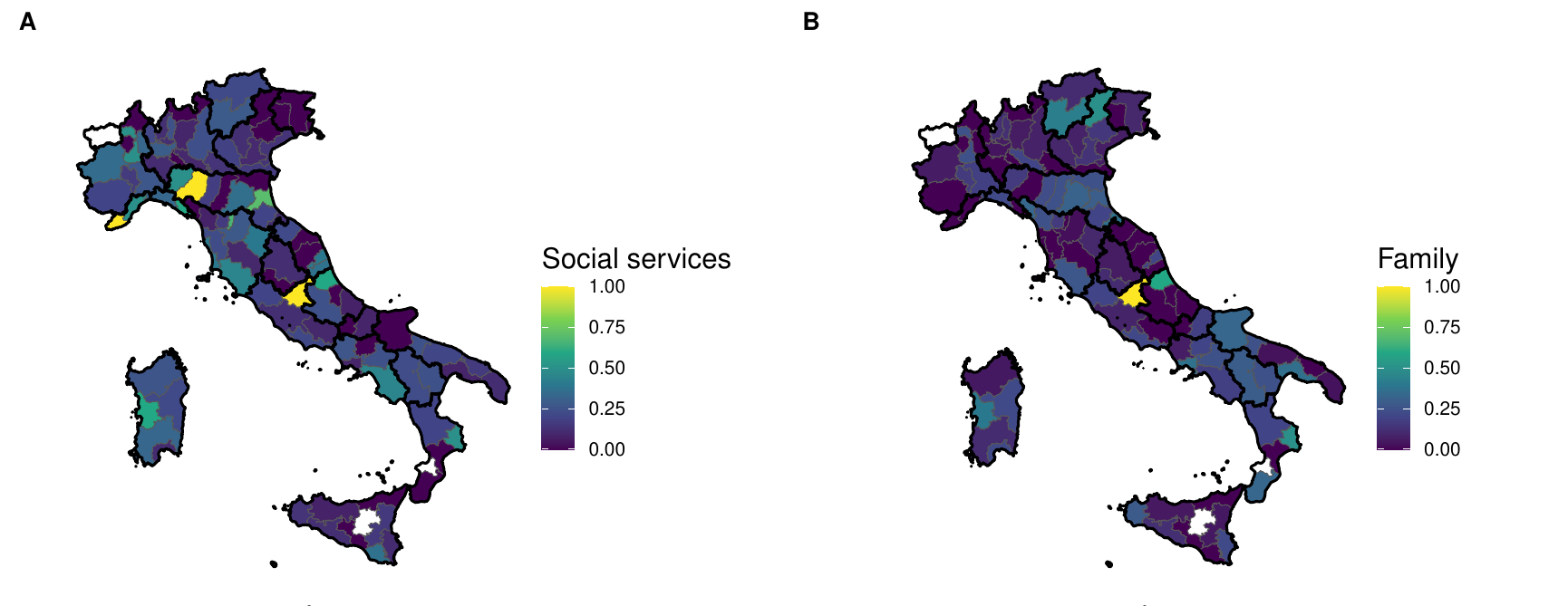}
			\caption{Provincial distribution of the \textit{Social services} (panel A) and \textit{Family} (panel B) indicators. Black lines denote regional boundaries. Darker shades indicate a value of 0, lighter shades a value of 1, and white denotes missing information.}
			\label{fig:RETTE_SERVSOCIALI_PREVISTO}
		\end{figure}

		Fee structures constitute a further determinant of inclusivity in Italian ECEC services. Fee reductions and exemptions are typically based on economic indicators and multiple criteria, including sibling enrollment, family residence, attendance type, disability status, social service recommendations, and other family-specific characteristics. At the national level, some form of fee reduction is available in $70.6\%$ of ECEC services. Full fee exemptions based on economic indicators are present in roughly one out of ten services overall, and in approximately one-fifth of public services, reaching nearly $16\%$ in Central Italy (ISTAT et al., \citeyear{Istat2022}). Greater uniformity in fee assistance is observed in large municipalities, where enrollment processes are centrally managed. These patterns reveal significant inequalities in access to ECEC services for families with similar socioeconomic conditions, particularly in regions where public provision is limited. \citetalias{Istat2022} stresses the importance of moving towards a more standardized and equitable fee reduction system to ensure consistent access across socioeconomic contexts. Figures~\ref{fig:PASTO}--\ref{fig:RETTE_SERVSOCIALI_PREVISTO} illustrate the provincial distribution of the economic-related indicators defined in Table~\ref{tab:var}.
		
			\section{Latent trait model}\label{app:ltm}
		\subsection{Latent trait model selection}
		
		We report model-selection analyses supporting the latent trait specification used to construct the inclusivity index. We compared four two-dimensional specifications by crossing two modeling choices: global versus multiple-group estimation and constrained versus unconstrained loading structure. The constrained specifications impose a simple structure in which social indicators load only on the first factor and economic indicators only on the second; the unconstrained specifications allow all items to load on both factors. In the multiple-group specifications, ownership defines the grouping variable. So, our final model is a configural multiple-group specification, i.e., the same loading structure is imposed across public and private services, and parameters can vary by group.
		
		Table \ref{tab:measurement_fit} reports the Root Mean Square Error of Approximation (RMSEA) with a 90\% confidence interval, the Tucker-Lewis Index (TLI), the Comparative Fit Index (CFI), and the Standardized Root Mean Square Residual (SRMSR) for the four candidate models. The multiple-group constrained specification has a lower RMSEA and satisfactory group-specific residual indices than the global constrained model, although its CFI and TLI do not improve. The unconstrained multiple-group model has the best approximate fit, as expected from its greater flexibility, but it does not preserve the theoretically defined separation between social and economic indicators. The final model was therefore not selected solely by maximizing empirical fit. The study prespecified the constrained structure from its substantive definition of inclusivity, providing dimensions with direct social and economic interpretation.

		\begin{table}[ht]
			\scriptsize
			\centering
			\caption{Fit statistics for the alternative two-dimensional latent trait models.}
			\label{tab:measurement_fit}
			\begin{tabular}{l|rrrrr}
				\toprule
				\textbf{Model} & \textbf{RMSEA} & \textbf{RMSEA 90\% CI} & \textbf{TLI} & \textbf{CFI} & \textbf{SRMSR} \\
				\midrule
				Global constrained & 0.074 & [0.066, 0.082] & 0.846 & 0.885 & 0.059 \\
				Multiple-group constrained & 0.045 & [0.039, 0.051] & 0.829 & 0.873 & 0.077 (public)/0.055 (private)\\
				Global unconstrained & 0.065 & [0.055, 0.074] & 0.882 & 0.934 & 0.05 \\
				Multiple-group unconstrained & 0.031 & [0.024, 0.039] & 0.917 & 0.954 & 0.073 (public)/0.041 (private) \\
				\hline
			\end{tabular}
		\end{table}
		
		We then assessed measurement invariance across ownership groups by comparing three nested specifications. The first is the proposed configural model. The second imposes metric invariance: discrimination parameters are constrained to be equal across public and private services, while latent variances are freely estimated. The third imposes scalar invariance: in addition to equal discrimination parameters, intercepts are constrained to be equal across groups, while latent means and variances are freely estimated. Table~\ref{tab:measurement_invariance} reports the results, indicating that full metric and scalar invariance across ownership groups are too restrictive.  The final model therefore retains the same theoretically defined organization of the social and economic dimensions while allowing item parameters to vary by ownership. However, we remark here that our model does not provide a common latent metric across groups; factor scores must then be interpreted as ownership-specific measures and are used only for within-group territorial analyses.

		\begin{table}[ht]
			\centering
			\begin{tabular}{l|rrrrrrrr}
				\hline
				\textbf{Model} & \textbf{Log-ln} & \textbf{AIC} & \textbf{BIC} & \textbf{RMSEA} & \textbf{TLI}& \textbf{CFI} & \textbf{LRT} & \textbf{df}\\
				\hline
				Configural & -6123.768 & 12339.54 & 12578.17  & 0.045 & 0.829  & 0.873 & -- & -- \\
				Metric & -6160.126 & 12396.25 & 12593.38  & 0.054 & 0.748 & 0.787 & 72.715 & 8 \\
				Scalar & -6280.196 & 12616.39 & 12761.65  & 0.076 & 0.502 & 0.524 & 240.141 & 10 \\
				\hline
			\end{tabular}
			\caption{Measurement invariance results across ownership groups.}
			\label{tab:measurement_invariance}
		\end{table}
		
		\subsection{Sampling-weight informativeness}
		
		We assess the sensitivity of the latent trait estimates to the use of sampling weights by re-estimating the model without incorporating weights in the likelihood.
		
\begin{figure}
			\centering
			\includegraphics[width=.4\columnwidth]{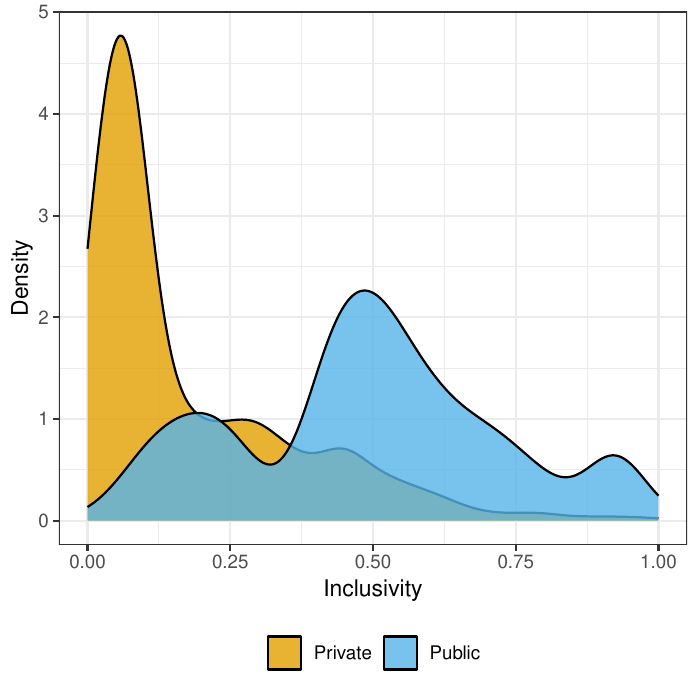}
			\caption{Density distributions of the scaled inclusivity index $\widetilde z$ defined in Equation \eqref{eq:scaledltm} for private (orange) and public (light blue) ECEC services without sampling weights. Scores are calibrated and rescaled separately within each ownership group; consequently, the horizontal positions and overlap of the two curves should not be interpreted as direct comparisons of absolute inclusivity levels.}
			\label{fig:hist_noweights}
		\end{figure}

		Figure~\ref{fig:hist_noweights} shows the density distribution of the scaled inclusivity index from the unweighted model, for comparison with Figure~\ref{fig:hist}. Visual inspection indicates that the overall distributional shape and dispersion of the index remain broadly similar under the two specifications.
		
\begin{figure}[!t]
			\centering
			\includegraphics[width=\columnwidth]{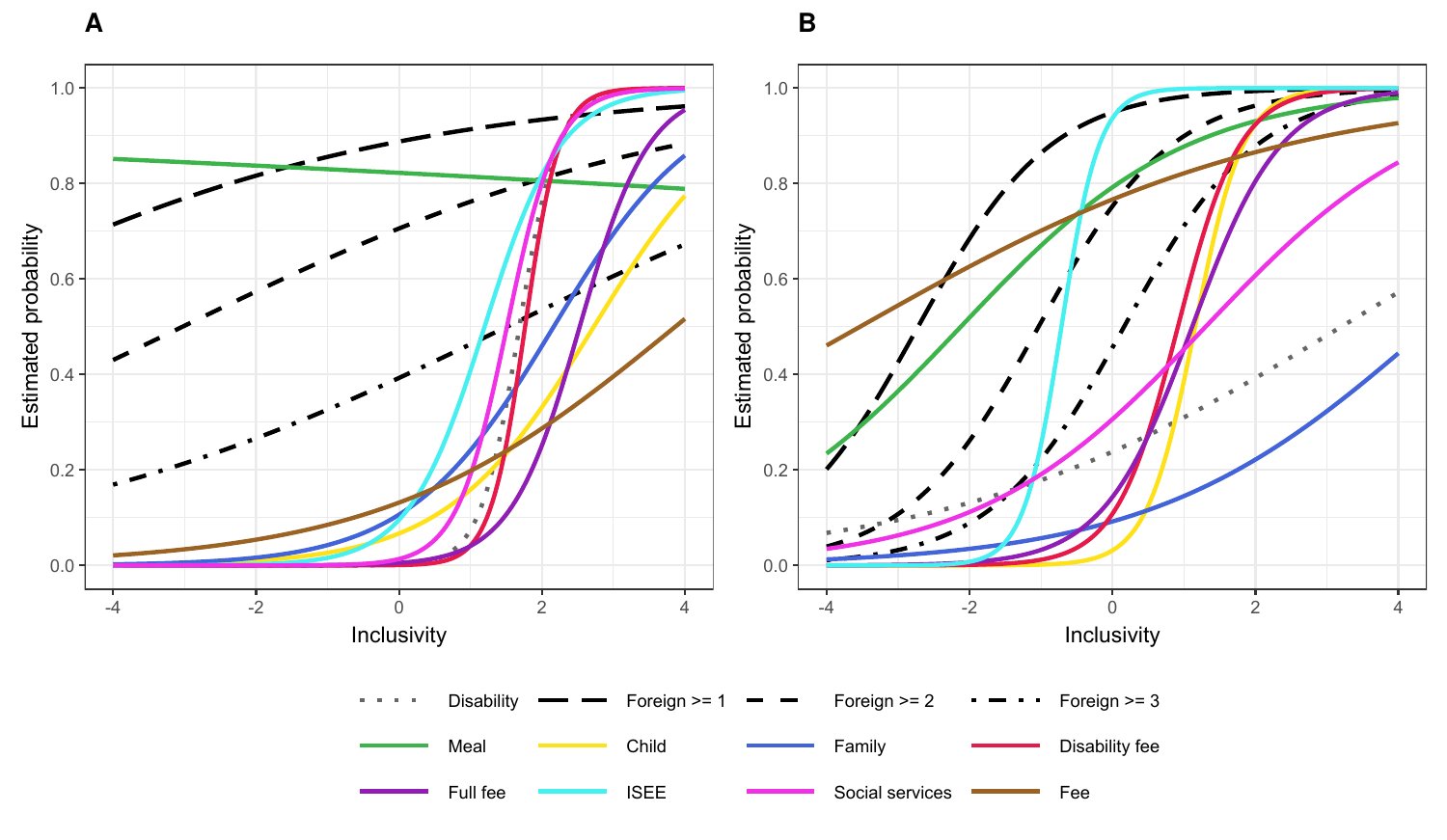}
			\caption{Estimated item-response functions from Equations \eqref{model}-\eqref{model_foreign} for private (panel A) and public (panel B) ECEC services without sampling weights.
				For binary indicators, the curves represent the estimated probability
				of a positive response. For the ordinal \textit{Foreign} item, the three
				black curves represent the estimated cumulative probabilities of
				attaining category $r$ or higher, for $r\in \{1,2,3\}$.}\label{fig:prob_noweight}
		\end{figure}

		Differences emerge more clearly at the level of individual item parameters. Figure~\ref{fig:prob_noweight} reports the estimated response probabilities from the unweighted model, corresponding to Figure~\ref{fig:prob_estimated}. The cumulative curves for \textit{Foreign} remain broadly similar under the two specifications in both ownership groups. For \textit{Disability}, the unweighted curve shifts toward lower values of the social dimension among private services, whereas it becomes flatter among public services. The \textit{Meal} indicator changes direction among private providers. Changes are also visible for \textit{Social services} and several fee-related items. In particular, among public services, the unweighted curve for \textit{Social services} is flatter, while that for \textit{Full fee} is steeper, changing their relative positions over part of the latent range.
		
		These differences suggest that sampling weights affect the magnitude of selected item parameters, particularly for indicators unevenly distributed across regions or ownership types. However, the overall dimensional structure and the distribution of the scaled inclusivity index remain largely stable.
		
		To formally assess the informativeness of the sampling weights, we applied the test proposed by \citet{asparouhov2007testing}, based on \citet{pfeffermann1993role}. Let $\boldsymbol{\theta} \in \mathbb{R}^{44}$ denote the vector of item intercepts and discrimination parameters in the latent trait model defined in Equations~\eqref{model}--\eqref{model_foreign}. Let $\boldsymbol{\theta}_w$ and $\boldsymbol{\theta}_0$ denote the item-parameter vectors targeted by the weighted and unweighted specifications, respectively, and let $\widehat{\boldsymbol{\theta}}_w$ and $\widehat{\boldsymbol{\theta}}_0$ denote their corresponding estimates. The null hypothesis is $H_0:\boldsymbol{\theta}_w=\boldsymbol{\theta}_0$. We then computed the statistic:
		$$
		Q=
		(\widehat{\boldsymbol{\theta}}_w-\widehat{\boldsymbol{\theta}}_0)^\top
		\left[
		\widehat{\operatorname{Var}}
		(\widehat{\boldsymbol{\theta}}_w-\widehat{\boldsymbol{\theta}}_0)
		\right]^{-1}
		(\widehat{\boldsymbol{\theta}}_w-\widehat{\boldsymbol{\theta}}_0),
		$$
		which, under $H_0$, is asymptotically distributed as a $\chi^2_q$ random variable, where $q$ is the number of item parameters included in the comparison \citep{asparouhov2007testing}. In our application, $q=44$. The global test rejected $H_0$ ($Q=185.394$,  $\mathrm{df}=44$, $p<0.001$). Thus, the test provides evidence that at least one item parameter differs between the weighted and unweighted specifications. To investigate whether this global rejection reflects a diffuse instability of the measurement model or localized parameter differences, we complemented the global test with marginal Wald-type statistics. For each parameter $\theta_j$, $j=1,\ldots,q$, we computed $Z_j=(\hat{\theta}_{w,j}-\hat{\theta}_{0,j})/\sqrt{\widehat{\mathrm{Var}}(\hat{\theta}_{w,j}-\hat{\theta}_{0,j})}$ and the associated two-sided $p$-values. The resulting marginal tests were adjusted for multiplicity using Holm's correction \citep{holm1979simple}. After adjustment, only a small subset of parameters remains significant at the 5\% level: the second cumulative intercept of the ordinal \textit{Foreign} item among private providers, the intercept of the \textit{Social Services} item among public providers, the discrimination parameter of the \textit{Full Fee} item among public providers, and the intercept of the \textit{Meal} item among private providers.

		\section{Small area estimation}
		\subsection{Diagnostic comparison of direct and small area estimates}
		\label{app:diagnosticsae-inclusivity}
		
\begin{figure}[htbp]
			\centering
			\includegraphics[width=0.8\textwidth]
			{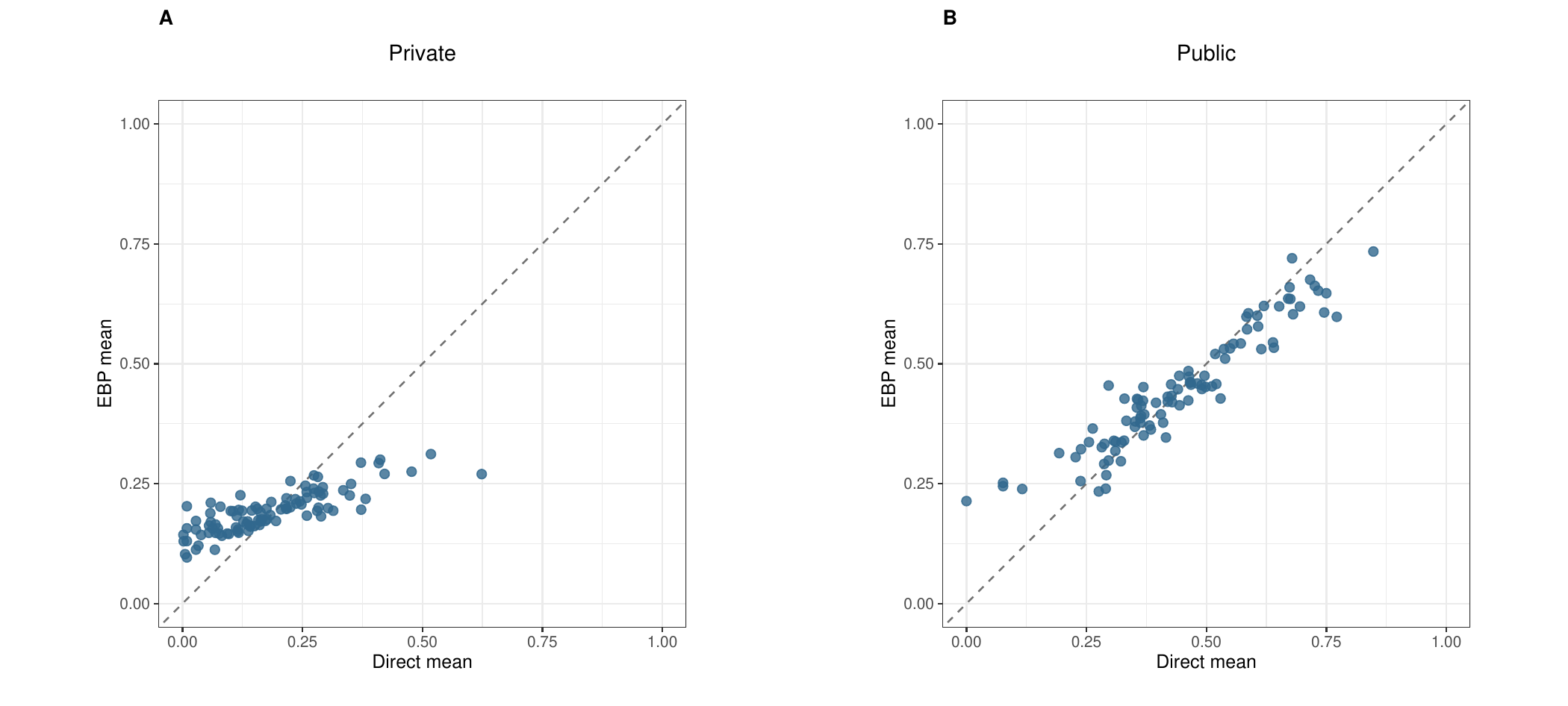}
			\caption{Provincial direct estimates ($x$-axis) and census SAE EBPs ($y$-axis) of the inclusivity index for private (panel A) and public (panel B) ECEC services. Each point represents a province; the dashed line indicates equality.}
			\label{fig:sae-direct-ebp}
		\end{figure}

\begin{figure}[htbp]
			\centering
			\includegraphics[width=0.8\textwidth]
			{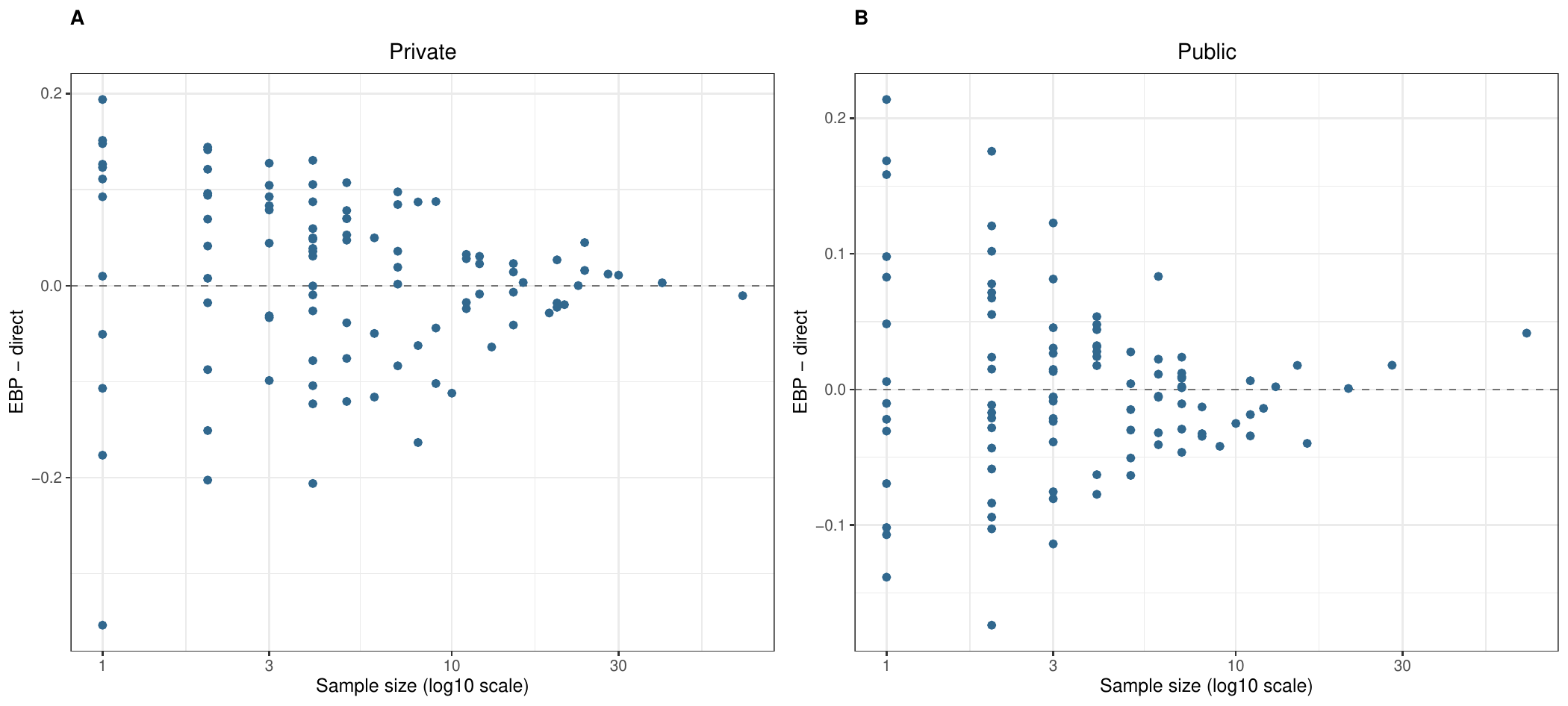}
			\caption{Differences between census SAE EBPs and direct estimates ($y$-axis) by provincial sample size ($x$-axis) for private (panel A) and public (panel B) ECEC services. Positive values indicate higher census SAE EBPs. Sample size is displayed on a logarithmic scale, and the dashed horizontal line indicates zero difference. The vertical scales differ between panels.}
			\label{fig:sae-differences}
		\end{figure}

		We compare provincial direct estimates with the census SAE EBPs, separately for private and public services. Figure~\ref{fig:sae-direct-ebp} displays the two estimates, while Figure~\ref{fig:sae-differences} plots their differences against the province-by-ownership sample size on a logarithmic scale. Comparisons are restricted to provinces with both estimates available: 99 for private services and 96 for public services.
		
		The estimates are positively associated in both groups, with stronger association for public services (Pearson's $r=0.943$; Spearman's $\rho=0.946$) than for private services ($r=0.829$; $\rho=0.814$). Among private services, census SAE EBPs occupy a smaller range than direct estimates, with upward adjustments at the lower end and downward adjustments at the upper end. Public-service estimates generally lie closer to the equality line. These patterns are consistent with model-based smoothing, while differences also reflect the use of register covariates and the distinct weighting of the direct and model-based estimators.
		
		For both ownership groups, differences are more dispersed in provinces with few sampled services and generally become smaller as sample size increases. This is consistent with the stronger influence of the model when local sample information is limited. 
		
		\subsection{Bootstrap estimation of the MSE of provincial inclusivity predictors}
		\label{app:mse-inclusivity}
		
		We describe the bootstrap procedure used to estimate the MSE of the provincial inclusivity predictors. Following \citet{moretti2021small}, we account for the uncertainty introduced by estimating the measurement model before performing small area prediction. The procedure of \citet{moretti2021small} is adapted to the latent trait model and the within-ownership normalization of the inclusivity index defined in Equations \eqref{model}--\eqref{eq:scaledltm}. It combines nonparametric resampling at the measurement stage with a parametric bootstrap under the Gaussian SAE working model. We use $B=500$ bootstrap replications.
		
		Let $w_k$ denote the final survey weight for service $k$. For each attempted replicate $b=1,\ldots,B$, the bootstrap algorithm performs the following steps. The superscript $\ast$ identifies quantities generated or re-estimated at the parametric SAE bootstrap stage.
		
		\begin{enumerate}
			\item \textbf{Resample services and re-estimate the latent trait model.} Draw $|\mathcal{K}_g|$ rows uniformly with replacement within each ownership group $g$. Items, covariates, province identifiers, and final survey
			weights are taken together with each selected row; selection probabilities are not proportional to $w_k$. Let $\mathcal{K}_g^{b}$ denote the resulting collection of resampled rows. Within this collection, $k$ indexes row occurrences, so repeated selections of the same service are retained as distinct observations. Fit the latent trait model \eqref{model}-\eqref{model_foreign} to the combined resampled data, using the final survey weights rescaled to the sample size as in the reference fit. Retain the correlated social and economic factors and align their signs with the reference loadings. For each resampled row, calculate the EAP scores, form the equally weighted composite score, and normalize it using the minimum and maximum within the corresponding resampled ownership group, following Equation~\eqref{eq:scaledltm}.
			
			\item \textbf{Estimate the bootstrap-generating SAE parameters.} Fit model~\eqref{eq:bhf} by REML to the resampled indices, separately for each ownership group, to obtain $\widehat{\boldsymbol\zeta}_g^{b}$, $\widehat\sigma_{s,g}^{2,b}$, and $\widehat\sigma_{\varepsilon,g}^{2,b}$. These estimates define the generating model for the parametric bootstrap within replicate $b$.
			
			\item \textbf{Generate provincial effects.} For every province-by-ownership domain in the register population, independently generate $ s_{pg}^{\ast, b}\sim N\!\left(0,\widehat\sigma_{s,g}^{2, b}\right)$. Use the same provincial effect for the artificial population and the artificial sample within the corresponding domain.
			
			\item \textbf{Generate individual errors.} Independently generate errors for every population unit and for
			every original sampled unit: $ \varepsilon_{kpg,\mathcal{U}}^{\ast, b}
			\sim N\!\left(0,\widehat\sigma_{\varepsilon,g}^{2, b}\right)$, $ \varepsilon_{kpg,\mathcal{K}}^{\ast,b}
			\sim N\!\left(0,\widehat\sigma_{\varepsilon,g}^{2, b}\right)$. Errors are independent across units, between the population and sample sets, and of the provincial effects. Thus, artificial sample and population responses share the provincial effect
			but not individual errors, even when records correspond to the same service. This construction does not require individual record linkage between the sample and the register population.
			
			\item \textbf{Calculate the true bootstrap provincial means.} For each $k\in \mathcal{U}_{pg}$, generate the artificial population response
			\begin{equation*}
				\widetilde z_{kpg,\mathcal{U}}^{\ast, b}
				=
				\mathbf r_{kpg}^{\top}\widehat{\boldsymbol\zeta}_g^{b}
				+s_{pg}^{\ast, b}
				+\varepsilon_{kpg,\mathcal{U}}^{\ast, b}.
			\end{equation*}
			The corresponding true bootstrap provincial mean is
			\begin{equation*}
				\mu_{pg}^{\ast, b}
				=
				\frac{1}{N_{pg}}
				\sum_{k\in \mathcal{U}_{pg}}\widetilde z_{kpg,\mathcal{U}}^{\ast,b}.
				\label{eq:mse-truth}
			\end{equation*}
			Artificial responses are generated on the original score scale under the Gaussian working model, without truncation or further normalization.
			
			\item \textbf{Generate the artificial sample responses.} Using the original sample covariates, set
			\begin{equation*}
				\widetilde z_{kpg,\mathcal{K}}^{\ast, b}
				=
				\mathbf r_{kpg}^{\top}\widehat{\boldsymbol\zeta}_g^{b}
				+s_{pg}^{\ast, b}
				+\varepsilon_{kpg,\mathcal{K}}^{\ast, b}, \qquad k \in \mathcal{K}_{pg}.
			\end{equation*}
			The original sample domain sizes and covariates are therefore held fixed at this stage. Retain the original final survey weights for computing the direct estimates.
			
			\item \textbf{Refit the SAE model and predict provincial means.} Fit the Gaussian working model by REML to the artificial sample, separately by ownership and without sampling weights, obtaining $\widehat{\boldsymbol\zeta}_g^{\ast, b}$ and
			$\widehat s_{pg}^{\ast, b}$. Calculate the analytical census SAE EBP
			\begin{equation*}
				\widehat\mu_{pg}^{\mathrm{EBP},\infty,\ast, b}
				=
				\overline{\mathbf r}_{\mathcal{U},pg}^{\top}
				\widehat{\boldsymbol\zeta}_g^{\ast, b}
				+
				\widehat s_{pg}^{\ast, b}.
			\end{equation*}
			For sampled domains, also calculate the direct estimator
			\begin{equation*}
				\widehat\mu_{pg}^{\mathrm{DIR}, \ast, b}
				=
				\frac{
					\sum_{k\in \mathcal{K}_{pg}}w_k\widetilde z_{kpg,\mathcal{K}}^{\ast, b}
				}{
					\sum_{k\in \mathcal{K}_{pg}}w_k
				}.
			\end{equation*}
			Set the predicted random effect to zero for domains without sampled services; their direct estimates are undefined. Record the squared prediction errors relative to $\mu_{pg}^{\ast, b}$, together with convergence information, warnings, and the occurrence of provincial predictions outside $[0,1]$.
			
			\item \textbf{Estimate MSE and relative uncertainty.} After $B$ attempted replicates, calculate
			\begin{equation*}
				\widehat{\operatorname{MSE}}_{\mathrm{boot}}
				\!\left(\widehat\mu_{pg}^{\mathrm{EBP},\infty}\right)
				=
				\frac{1}{B}\sum_{b=1}^{B}
				\left(
				\widehat\mu_{pg}^{\mathrm{EBP},\infty,\ast, b}
				-
				\mu_{pg}^{\ast, b}
				\right)^2.
			\end{equation*}
			Apply the same expression to the direct estimator, where defined, using its corresponding bootstrap predictions. Replicates that fail are not automatically replaced. Under the implementation's completeness rule, an MSE is reported as missing if any of its $B$ required squared errors is missing or invalid. The RRMSE is
			\begin{equation*}
				\operatorname{RRMSE}_{pg}(\%)
				=
				100\,
				\frac{
					\sqrt{
						\widehat{\operatorname{MSE}}_{\mathrm{boot}}
						\!\left(\widehat\mu_{pg}^{\mathrm{EBP},\infty}\right)
					}
				}{
					\left|\widehat\mu_{pg}^{\mathrm{EBP},\infty}\right|
				}.
			\end{equation*}
			The RRMSE is reported as missing when the original prediction is zero or unavailable.
			
		\end{enumerate}

			\section{Random-effects $\tau$-quantile regression}
		\label{app:lqmm_mqmm_comparison}

		As a robustness check, we compared the results obtained from the weighted LQMM \citep{geraci2007quantile,geraci2014linear}, used in the main analysis, with those obtained from the weighted $M$-quantile mixed model proposed by \citet{spagnolo2020use}. Both approaches were applied separately to private and public ECEC services at $\tau\in\{0.25,0.50,0.75\}$, using the same regional clustering structure and ownership-specific regional weights. Because quantile and $M$-quantile models target related but distinct distributional summaries, numerical equality between their estimates is not expected. We therefore assessed robustness in terms of the relative regional ordering produced by the two approaches. For each ownership group and value of $\tau$, regions were ranked from 1, corresponding to the lowest estimated value, to 19, corresponding to the highest.
		
\begin{figure}[ht]
			\centering
			\includegraphics[width=.8\linewidth]{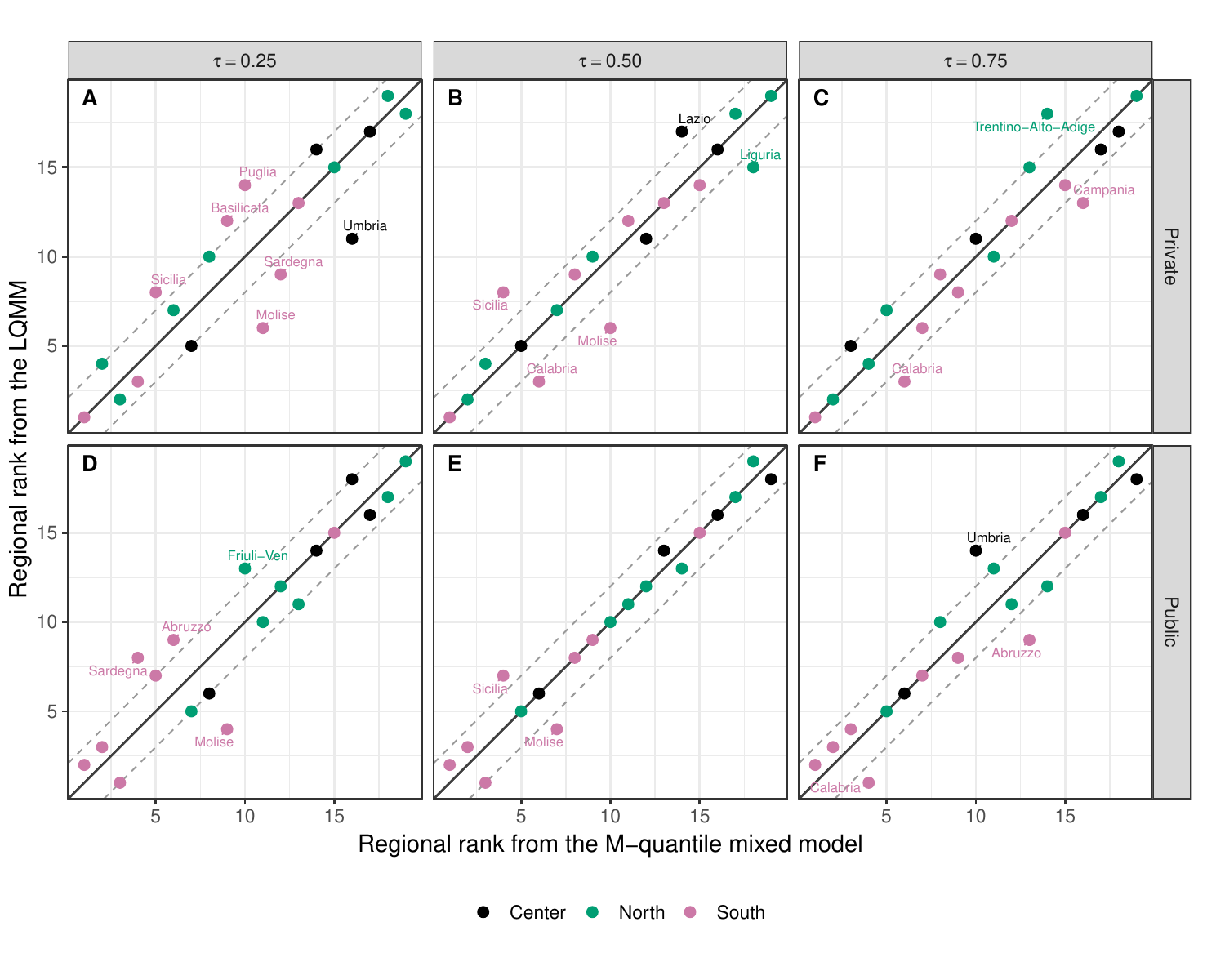}
			\caption{Comparison of regional ranks obtained from the weighted $M$-quantile mixed model \citep{spagnolo2020use} ($x$-axis) and the LQMM \citep{geraci2007quantile,geraci2014linear} ($y$-axis), separately by ownership group and for $\tau\in\{0.25,0.50,0.75\}$. Ranks are computed within each ownership--$\tau$ combination, with 1 denoting the lowest estimated value and 19 the highest. The solid diagonal line represents exact rank agreement, while the dashed lines delimit differences of no more than two rank positions. Points represent regions, colors identify macro-areas, and region labels highlight the largest rank discrepancies.}
			\label{fig:lqmm_mqmm}
		\end{figure}

		Figure~\ref{fig:lqmm_mqmm} shows broad agreement in the regional ordering obtained from the two methods. Across the six ownership-by-$\tau$ comparisons, Spearman's rank correlation ranged from $0.900$ to $0.975$, and the mean absolute rank difference ranged from $0.74$ to $1.89$ positions; between $68.4\%$ and $89.5\%$ of regions differed by no more than two positions. Agreement is closest at $\tau=0.50$, particularly among public services, whereas rank differences are more evident at the lower and upper ends of the distribution. These differences mainly concern regions occupying intermediate positions, while regions located near the lower and upper ends generally retain similar relative positions under the two approaches. Thus, the comparison supports the robustness of the general territorial patterns identified by the LQMM.
		
	\clearpage
	\bibliographystyle{apalike}
	\bibliography{reference}

\end{document}